\documentclass[11pt]{article}
\title{The Interaction of Dirac Particles with Non-Abelian Gauge Fields
and Gravity -- Bound States}
\author{Felix Finster, Joel Smoller\thanks{Research supported in part by the
NSF, Grant No.\ DMS-G-9802370.}, and Shing-Tung
Yau\thanks{Research supported in part 
by the NSF, Grant No.\ 33-585-7510-2-30.}}
\date{January 2000}

\newcommand{\spc}{\;\;\;\;\;\;\;\;\;\;}
\newcommand{\bra}{\mbox{$< \!\!$ \nolinebreak}}
\newcommand{\ket}{\mbox{\nolinebreak $>$}}
\newcommand{\C}{\:\mbox{\rm I \hspace{-1.25 em} {\bf C}}}
\newcommand{\R}{\mbox{\rm I \hspace{-.8 em} R}}
\newcommand{\1}{\mbox{\rm 1 \hspace{-1.05 em} 1}}

\newcommand{\Tr}{\mbox{Tr\/}}

\begin{document}
\include{epsf}
\maketitle

\begin{abstract}
We consider a spherically symmetric, static system of a Dirac particle
interacting with classical gravity and an $SU(2)$ Yang-Mills
field. The corresponding Einstein-Dirac-Yang/Mills equations are derived.
Using numerical methods, we find different types of soliton-like solutions of
these equations and discuss their properties. Some of these solutions 
are stable even for arbitrarily weak gravitational coupling.
\end{abstract}

\section{Introduction}
\setcounter{equation}{0}
The coupling of gravity to other classical force fields and to quantum
mechanical particles has led to many interesting solutions of Einstein's
equations and has given some insight into the nature of the nonlinear
interactions. The first such examples are the Bartnik-McKinnon (BM)
solutions of the Einstein-Yang/Mills (EYM) equations~\cite{BM}. For 
these solutions, the repulsive Yang-Mills force compensates the
attractive gravitational force; unfortunately, these solutions are
unstable~\cite{SZ}. If, on the other hand, one considers quantum mechanical
Dirac particles, the gravitational attraction  is balanced by the 
repulsion due to the Heisenberg Uncertainty Principle, and this leads to
stable bound states of the resulting Einstein-Dirac (ED) system~\cite{FSY1}.
However, a pure ED system is somewhat artificial, because physical Dirac
particles also exhibit electroweak and strong interactions, which
in all realistic situations are much stronger than gravity. Thus the
question arises if Dirac particles in a gravitational field
still form bound states if an additional strong coupling to a non-abelian
Yang-Mills (YM) field is taken into account (the
case of an abelian gauge field was considered in~\cite{FSY2}). Related
questions are, do the BM solutions become stable if one includes Dirac
particles, and which physical effects does the nonlinear coupling in the
Einstein-Dirac-Yang/Mills (EDYM) equations lead to?

In order to address these questions, we consider here a static, spherically
symmetric EDYM system of one Dirac particle in a gravitational field 
and an $SU(2)$ Yang-Mills field. In this system, the spinors couple 
only to the magnetic component of the YM field, and we thus obtain a 
consistent ansatz by setting the electric component identically equal to 
zero. In the limit of weak coupling of the spinors, our
system goes over to the EYM system~\cite{BM}. In contrast to
the two-particle singlet state studied in~\cite{FSY1}, we consider here only
one Dirac particle (this becomes possible because the inclusion of the
$SU(2)$ Yang-Mills field changes the representation of the rotation group on
the spinors; see Section~\ref{sec2}).
Thus one cannot recover exactly the ED system~\cite{FSY1},
but the limit of a weak Yang-Mills field yields equations which are closely
related to the ED equations of the two-particle singlet.

By numerically seeking bound states of the EDYM system, we find a
surprisingly rich solution structure. First of all, we find solutions where
the Dirac
particle is bound by the gravitational attraction, and where the Dirac
particle also generates a YM field. Stable solutions of this type exist also
in the physically realistic situation of weak gravitational and
strong YM coupling. This result shows that the magnetic
component of the YM field, which usually has a repulsive effect (like e.g.\
for the BM solutions) cannot prevent Dirac particles from forming stable bound
states. We also find other types of solutions where the binding comes about
through the nonlinear interaction of all three fields.
These solutions have stable and unstable branches, whereby the stable 
solutions are found for weak gravitational coupling, provided that 
the YM coupling is sufficiently strong, but not too strong. Finally, we 
study the relation between these solutions and the BM 
solutions. We find one-parameter families of solutions which join the 
BM ground state with our new solutions. This shows that the BM ground 
state can be made stable by the inclusion of a Dirac particle, but 
only if the coupling to the spinors is sufficiently strong. The first 
excited BM state, on the other hand, cannot be joined with our new 
solutions. Namely, perturbing this state by an additional Dirac 
particle yields a separate unstable branch of EDYM solutions.

The plan of the paper is as follows. In Section~\ref{sec2} we derive the
$SU(2)$-EDYM equations. In Section~\ref{sec3} we obtain a
limiting system constructed by letting the gravitational constant tend
to zero and letting the YM coupling constant tend to infinity. We find
numerical solutions of this system and discuss their properties.
In the last section, we consider solutions of the full EDYM equations,
obtained by tracing the solutions of our limiting system and the BM
solutions while continuously varying the coupling constants.

\section{Derivation of the Equations}
\setcounter{equation}{0}
\label{sec2}
The general EDYM equations are obtained by variation over Lorentzian
metrics $g_{ij}$, YM connections ${\cal{A}}$, and Dirac wave functions $\Psi$,
of the action
\begin{equation}
S \;=\; \int \left( \frac{1}{16 \pi \:\kappa} \:R \:+\:
\overline{\Psi} (G-m) \Psi \:-\: \frac{1}{16 \pi\:e^2} \:\Tr (F_{ij} \:
F^{ij}) \right) \sqrt{-\det g} \:d^4x \;\;\;,
\label{eq:1A}
\end{equation}
where $R$ is scalar curvature, $G$ is the Dirac operator (which depends
on the gravitational and YM fields), $F_{ij}$ is the YM field tensor,
and where the trace is taken over the YM index.
The gravitational and YM coupling constants are denoted by $\kappa$ and
$e$, respectively. The appearance of the factor $e^{-2}$ in (\ref{eq:1A})
requires a brief explanation. In contrast to the usual form of the
gauge-covariant derivative $D_j=\partial_j-ie A_j$, we use here the
convention $D_j=\partial_j-iA_j$ (this makes it possible to work with
the particularly convenient form of the gauge potentials used 
in~\cite{BM}). Our convention is obtained from the usual one by the
transformation $A_j \rightarrow e^{-1} \:A_j$. Under this transformation,
the field strength tensor behaves like $F_{ij} \rightarrow e^{-1} \:F_{ij}$,
and this gives rise to the factor $e^{-2}$ in~(\ref{eq:1A}).

In this paper, we shall study a particular EDYM system, which is 
obtained as follows. First of all, we consider a spherically symmetric,
static metric in polar coordinates,
\begin{equation}
ds^2 \;=\; \frac{1}{T(r)^2} \:dt^2 \:-\: \frac{1}{A(r)} \:dr^2 \:-\:
r^2 \:d\vartheta^2 \:-\: r^2\: \sin^2 \vartheta \:d \varphi^2 \;\;\;,
\label{eq:1a}
\end{equation}
with positive functions $A$ and $T$. The Einstein tensor corresponding
to this metric is given in~\cite{FSY1}. Moreover, as in~\cite{BM}, we
restrict attention to the magnetic component of an $SU(2)$ Yang-Mills field
and choose for the YM potential the ansatz
\begin{equation}
{\cal{A}} \;=\; w(r) \:\tau^1\: d\vartheta \:+\: (\cos \vartheta \:
\tau^3 \:+\: w(r) \:\sin \vartheta \:\tau^2) \:d\varphi
\label{eq:1b}
\end{equation}
with a real function $w$, where $\vec{\tau}=\frac{1}{2} \: \vec{\sigma}$ is
the standard basis of $su(2)$ ($\vec{\sigma}$ are the Pauli matrices). The
YM current $j$ and energy-momentum tensor
$T^i_j =\frac{1}{4 \pi e^2} \:{\mbox{Tr }} (F^{ia} F_{ja} - \frac{1}{4}
\:F^{ab} F_{ab} \:\delta^i_j)$ are computed to be
\begin{eqnarray}
j &=& \frac{1}{4 \pi e^2}
\left(-\frac{A}{2 r^2} \:w'' \:-\: \frac{A' T
- 2 A T'}{4 r^2 \:T} \:w' \:-\: \frac{w (1-w^2)}{2 r^4} \right)
\left(\sigma^1 \:\frac{\partial}{\partial \vartheta} \:+\: \sigma^2 \:
\csc (\vartheta) \: \frac{\partial}{\partial \varphi} \right) \nonumber \\
T^0_0 &=& \frac{1}{4 \pi e^2} \left(-\frac{2}{r^4} \:(1-w^2)^2
\:-\: \frac{4}{r^2} \:A\:w'^2 \right) \label{eq:TT0} \\
T^1_1 &=& \frac{1}{4 \pi e^2} \left(-\frac{2}{r^4} \:(1-w^2)^2
\:+\: \frac{4}{r^2} \:A\:w'^2 \right) \\
T^2_2 &=& T^3_3 \;=\; \frac{1}{4 \pi e^2} \left(
\frac{2}{r^4} \:(1-w^2)^2 \right) \;\;\;, \label{eq:TT1}
\end{eqnarray}
and all other components vanish.
When coupled to the Dirac spinors, the YM potential (\ref{eq:1b}) has the
disadvantage that it depends on $\vartheta$ and $\varphi$, in a way 
which makes it difficult to separate variables in the Dirac equation.
To remedy this, we perform the $SU(2)$ gauge transformation
${\cal{A}}_j \rightarrow U {\cal{A}}_j U^{-1} + i U (\partial_j U^{-1})$ with
\[ U(\vartheta, \varphi) \;=\; \exp \left(-i \varphi \:\tau^1 \right) \:
\exp \left(-i(\vartheta+\pi) \:\tau^3 \right) \:
\exp \left( \frac{i \pi}{2} \:\tau^2 \right) \;\;\;. \]
The resulting YM potential is
\begin{equation}
{\cal{A}} \;=\; (w-1) \:r\: \sin \vartheta \:(\tau^\varphi \:d\vartheta
\:-\: \tau^\vartheta \:d\varphi) \;\;\;,
\label{eq:1c}
\end{equation}
where we use the following ``polar'' linear combinations of the $\tau$
matrices,
\begin{eqnarray}
\tau^r &=& \tau^1 \:\cos \vartheta \:+\: \tau^2 \:
	\sin \vartheta \: \cos \varphi \:+\: \tau^3 \: \sin \vartheta \:
	\sin \varphi \nonumber \\
\tau^\vartheta &=& \frac{1}{r} \left( -\tau^1 \:\sin \vartheta \:+\: \tau^2 \:
	\cos \vartheta \: \cos \varphi \:+\: \tau^3 \: \cos \vartheta \:
	\sin \varphi \right) \nonumber \\
\tau^\varphi &=& \frac{1}{r \sin \vartheta} \left( -\tau^2 \: \sin \varphi
	\:+\: \tau^3 \: \cos \varphi \right) \;\;\; . \label{eq:1d}
\end{eqnarray}
By minimally coupling the $SU(2)$ potential (\ref{eq:1c}) to the Dirac operator
in the gravitational field~\cite[Eq.\ (2.23)]{FSY1}, we obtain the 
Dirac operator
\begin{eqnarray}
G &=& i T \:\gamma^t \partial_t \:+\: \gamma^r \left(i \sqrt{A} 
\partial_r + \frac{i}{r} \:(\sqrt{A}-1) -\frac{i}{2} \:\sqrt{A} 
\:\frac{T'}{T} \right) \:+\: i \gamma^\vartheta 
\partial_\vartheta \:+\: i \gamma^\varphi \partial_\varphi \nonumber \\
&&\:+\: \frac{2i}{r} \:(w-1) \:(\vec{\gamma} \vec{\tau}
- \gamma^r \tau^r) \:\tau^r
\;\;\;,\;\;\;\;\: \label{eq:1e}
\end{eqnarray}
where $(\gamma^j)_{j=t,r,\vartheta, \varphi}$ are, in analogy to (\ref{eq:1d}),
the Dirac matrices of Minkowski space in polar coordinates, where we 
work in the Dirac representation,
\begin{equation}
	\gamma^t \;=\; \left( \begin{array}{cc} \1 & 0 \\ 0 & -\1 
	\end{array} \right) \;\;\;,\spc \vec{\gamma} \;=\;
	\left( \begin{array}{cc} 0 & \vec{\sigma} \\ -\vec{\sigma} & 0
	\end{array} \right) \;\;\;.
\end{equation}
Notice that the Dirac operator (\ref{eq:1e}) acts on 8-component wave
functions; this is because the additional YM index doubles the number of
components compared to usual Dirac spinors. More precisely, it is convenient
to regard the wave functions as sections of
\begin{equation}
\C^8 \;=\; \C^2_{\mbox{\scriptsize{up/down}}} \otimes
\C^2_{\mbox{\scriptsize{large/small}}} \otimes
\C^2_{\mbox{\scriptsize{YM}}} \;\;\;, \label{eq:2fn}
\end{equation}
where $\C^2_{\mbox{\scriptsize{up/down}}}$ describes the two spin orientations,
$\C^2_{\mbox{\scriptsize{large/small}}}$ corresponds to the upper and lower
components of the Dirac spinor (i.e., usual Dirac spinors are sections of
$\C^4 = \C^2_{\mbox{\scriptsize{up/down}}} \otimes
\C^2_{\mbox{\scriptsize{large/small}}}$), and
$\C^2_{\mbox{\scriptsize{YM}}}$ is acted upon by the $SU(2)$ gauge group.
For clarity, we shall refer to the factors in (\ref{eq:2fn}) by separate
indices, i.e.\ we write a wave function $\Psi$ as
$(\Psi^{\alpha u a})_{\alpha, u, a=1,2}$, where $\alpha$, $u$, and $a$
label the components of $\C^2_{\mbox{\scriptsize{up/down}}}$,
$\C^2_{\mbox{\scriptsize{large/small}}}$, and
$\C^2_{\mbox{\scriptsize{YM}}}$, respectively. Thus the operators
$\vec{\tau}$ act on the index $a$,
the spin operators $\vec{S}$ are given by
$\vec{S} = \frac{1}{2}\:\vec{\sigma}$ acting on the Greek indices,
and $\gamma^t$ coincides with the matrix $\gamma^t={\mbox{diag}}(1,-1)$
acting on the index $u$, i.e.\
\[ \gamma^t \:\Psi^{\alpha u a} \;=\; \left\{ \begin{array}{cl}
\Psi^{\alpha u a} & {\mbox{if $u=1$}} \\
-\Psi^{\alpha u a} & {\mbox{if $u=2$}} \end{array} \right. \;\;\;. \]
It is apparent in (\ref{eq:1e}) that the Dirac operator commutes with
the three operators
\begin{equation}
\vec{J} \;=\; \vec{L} + \vec{S} + \vec{\tau} \;\;\;,
\label{eq:1f}
\end{equation}
where $\vec{L}$ is angular momentum. It is very convenient to regard the
operators $\vec{J}$ as the total angular momentum operators of the system.
Since the total angular momentum operators are the infinitesimal generators of
rotations (as explained for angular momentum in \cite[par.\ 26]{LL}), we can
then interpret (\ref{eq:1f}) as saying that the inclusion
of the YM field influences the representation of the rotation group on the
spinors. The Dirac operator is invariant under this group representation,
because the operators $\vec{J}$ commute with $G$; this makes
spherical symmetry of the Dirac operator manifest.

Since (\ref{eq:1f}) coincides with the formula for the addition of angular
momentum $\vec{L}$ to two spin-$\frac{1}{2}$-operators $\vec{S}$ and
$\vec{\tau}$, it is clear that the operators $\vec{J}$ have integer
eigenvalues.
Thus we can expect that the operator $J^2$ has a nontrivial kernel.
In this case, the simplest spherically symmetric ansatz for the Dirac particles
would be to take one Dirac particle whose wave function is in the kernel of
$J^2$. We now work out this ansatz in detail, whereby we consider $\vec{J}$ as
operators on the spinors $\Phi^{\alpha a}(\vartheta, \varphi)$ on $S^2$
(i.e.\ the $\Phi^{\alpha a}$ are sections of
$\C^4 = \C^2_{\mbox{\scriptsize{up/down}}} \otimes
\C^2_{\mbox{\scriptsize{YM}}}$). Adding the two spin operators
$\vec{S}$ and $\vec{\tau}$, we can decompose
$\C^2_{\mbox{\scriptsize{up/down}}} \otimes \C^2_{\mbox{\scriptsize{YM}}}$
into the direct sum of one state of total spin zero and three states of
total spin one (see \cite[par.\ 31]{LL}; these states are usually called
the singlet and triplet states, respectively).
By subsequently adding the angular momentum $\vec{L}$ according to the standard
rules for the addition of angular momentum \cite[par.\ 31]{LL}, one sees
that the operator $J^2$ has indeed a nontrivial kernel.
More precisely, the kernel of
$J^2$ is two-dimensional, spanned by two vectors $\Phi_0$ and
$\Phi_1$ with angular momentum zero and one, respectively. The state $\Phi_0$
is (up to a phase) uniquely characterized by the conditions
\begin{equation}
\vec{L} \:\Phi_0 \;=\; 0 \;=\; (\vec{S} + \vec{\tau}) \:\Phi_0
\spc {\mbox{and}} \spc \| \Phi_0 \|_{S^2} \;=\; 1 \;\;\;.
\label{eq:1g}
\end{equation}
Using (\ref{eq:1g}), we can write $\Phi_1$ as
\begin{equation}
\Phi_1 \;=\; 2 S^r \:\Phi_0 \;=\; -2 \tau^r \:\Phi_0 \;\;\;.
\label{eq:1i}
\end{equation}
Namely, representing $S^r$ and $\tau^r$ in the form
\[ S^r \;=\; \vec{x} \vec{s} \spc {\mbox{and}} \spc \tau^r \;=\; \vec{x}
\vec{\tau} \;\;\;, \]
and using the standard commutation relations between the components
of $\vec{L}$, $\vec{x}$, and $\vec{S}$ (see \cite[pars.\ 26 and 54]{LL}),
we obtain that
\begin{eqnarray}
\vec{J} \:\Phi_1 &=& 2 \:[\vec{J},\:S^r] \:\Phi_0
\;=\; 2 \:[\vec{L},\:(\vec{x} \vec{S})]\:\Phi_0
\:+\: 2 \:[\vec{S}+\vec{\tau},\:(\vec{x} \vec{S})]\:\Phi_0 \nonumber \\
&=& -2i \:\vec{x} \wedge \vec{S} \:\Phi_0 \:+\: 2i \:\vec{x} \wedge \vec{S}
\:\Phi_0 \;=\; 0 \label{eq:1k}
\end{eqnarray}
(where $\wedge$ is the wedge product in $\R^3$), and
\[ \| \Phi_1 \|_{S^2}^2 \;=\; \int_{S^2} \bra 2 S^r \Phi_0, \:2 S^r 
\Phi_0 \ket \:d\omega \;=\; \| \Phi_0 \|_{S^2}^2 \;=\; 1 \;\;\;. \]
One can verify directly that $\Phi_1$ has angular momentum one; namely
\begin{eqnarray*}
L^2 \:\Phi_1 &=& 2 L^2 \:S^r\:\Phi_0 \;=\; 2 \vec{L} \:[\vec{L},\:S^r]
\:\Phi_0 \;=\; -2i \:\vec{L} (\vec{x} \wedge \vec{S}) \:\Phi_0 \\
&=& -2i \:[\vec{L},\:(\vec{x} \wedge \vec{S})]\:\Phi_0 \;=\;
4 \:(\vec{x} \vec{S}) \:\Phi_0 \;=\; l(l+1) \:\Phi_1
\end{eqnarray*}
with $l=1$. Furthermore, using the fact that $(2S^r)^2=1=(2 \tau^r)^2$ and
$S^2=\frac{3}{4}=\tau^2 $, we obtain that
\begin{eqnarray}
\Phi_0 &=& 2 S^r \:\Phi_1 \;=\; -2 \tau^r \:\Phi_1 \label{eq:1l} \\
\vec{S} \vec{\tau} \:\Phi_0 &\stackrel{(\ref{eq:1g})}{=}& -\vec{S} \vec{S}\:
\Phi_0 \;=\; -\frac{3}{4} \:\Phi_0 \label{eq:1m} \\
\vec{S} \vec{\tau} \:\Phi_1 &=& \frac{1}{2} \:((\vec{S}+\vec{\tau})^2 - S^2
- \tau^2) \:\Phi_1 \;\stackrel{(\ref{eq:1k})}{=}\; 
\frac{1}{2} \:(L^2 - S^2 - \tau^2) \:\Phi_1 \;=\;
\frac{1}{4} \:\Phi_1 \label{eq:1n} \\
\vec{S} \vec{L} \:\Phi_0 &=& 0 \\
\vec{S} \vec{L} \:\Phi_1 &=& \frac{1}{2} \:((\vec{S}+\vec{L})^2 - 
S^2 - L^2)\:\Phi_1 \;=\; \frac{1}{2}\: (\tau^2 - S^2 - L^2)\:\Phi_1
\;=\; -\Phi_1 \;\;\;. \;\;\;\;\;\; 
\end{eqnarray}
Finally, it is useful to observe that (cf.\ \cite[equation (3.3)]{FSY3})
\begin{equation}
S^\vartheta \partial_\vartheta + S^\varphi \partial_\varphi \;=\;
-\frac{2}{r} \:S^r \:(\vec{S} \vec{L}) \;\;\;.
\label{eq:1o}
\end{equation}
Using the relations (\ref{eq:1g})--(\ref{eq:1o}), one can easily compute
the Dirac operator (\ref{eq:1e}) on the invariant subspace $J^2=0$. It turns
out that we obtain a consistent ansatz for the Dirac wave function
by setting
\begin{equation}
\Psi^{\alpha u a}(t,r,\vartheta,\varphi)
\;=\; e^{-i \omega t} \:\frac{\sqrt{T(r)}}{r} \left(
\alpha(r) \:\Phi_0^{\alpha a}(\vartheta, \varphi) \:\delta_{u,1}
\:+\: i \beta(r) \: \Phi_1^{\alpha a}(\vartheta, \varphi) \:\delta_{u,2}
\right)
\label{eq:1p}
\end{equation}
with real functions $\alpha$ and $\beta$, where $\omega>0$ is the energy of
the Dirac particle, and $\delta_{.,.}$ is the Kronecker delta.
For this ansatz, the Dirac equation
reduces to the system of ODEs
\begin{eqnarray}
\sqrt{A} \:\alpha' &=& \frac{w}{r} \:\alpha \:-\: (m + \omega T) \:
\beta \label{eq:1d1} \\
\sqrt{A} \:\beta' &=& (-m+\omega T) \:\alpha \:-\: \frac{w}{r} \:\beta
\;\;\;. \label{eq:1d2}
\end{eqnarray}
The Dirac current $j$ and Dirac energy-momentum tensor $T_{jk} = {\mbox{Re }}
(\overline{\Psi} G_{(j} D_{k)} \Psi)$ corresponding to the ansatz (\ref{eq:1p})
are obtained by a straightforward computation similar to that in~\cite{FSY1}.
The result is
\begin{eqnarray*}
j &=& -\frac{2T}{r^3} \: \alpha \beta 
\left( \sigma^1 \:\frac{\partial}{\partial \vartheta} \:+\: \sigma^2 \:
\csc (\vartheta) \:\frac{\partial}{\partial \varphi} \right) \\
T^0_0 &=& \frac{\omega T^2}{r^2} \:(\alpha^2+\beta^2) \\
T^1_1 &=& -\frac{\omega T^2}{r^2} \:(\alpha^2+\beta^2) \:+\: 2\:
\frac{T}{r^3} \:w\:\alpha \beta \:+\: \frac{mT}{r^2} \:(\alpha^2 - \beta^2) \\
T^2_2 &=& T^3_3 \;=\; -\frac{T}{r^3} \:w\:\alpha \beta \;\;\;,
\end{eqnarray*}
and all other components vanish.
The normalization condition for the spinors is (as in~\cite{FSY1}),
\begin{equation}
\int_0^\infty (\alpha^2+\beta^2) \:\frac{T}{\sqrt{A}} \;=\; 1 \;\;\;.
\label{eq:1nn}
\end{equation}

By substituting the formulas for the YM current and energy-momentum
tensor into the Einstein and YM equations, we obtain the following
system of ODEs,
\begin{eqnarray}
r \:A'
&=& 1-A \:-\:\frac{\kappa}{e^2}\:\frac{{{(1-{w^2})}^2}}{{r^2}}
\:-\:2 \kappa\:\omega  {T^2} \big({{\alpha }^2}+ {{\beta }^2}\big)
\:-\:\frac{2 \kappa}{e^2} \:A \:{w'^2}   \label{eq:1r} \\
2 r A \:\frac{T'}{T} &=& -1+A \:+\:
\frac{\kappa}{e^2} \:\frac{{{(1-{w^2})}^2}}
{{r^2}} \:+\:2 \kappa \:m T \big({{\alpha }^2}-{{\beta }^2}\big)
\:-\:2 \kappa \:\omega  \:{T^2} \big({{\alpha }^2}+{{\beta }^2}\big)
\nonumber \\
&&+4 \kappa \:\frac{T}{r} \:w\:\alpha \beta
\:-\:\frac{2 \:\kappa}{e^2} \:A\: w'^2 \label{eq:1s} \\
{r^2} A \:w'' &=&
-\big(1-{w^2}\big) \:w\:+\:e^2\:r T \alpha  \beta \:-\: {r^2}
\:\frac{A' \:T - 2 A \:T'}{2T} \:w' \;\;\;.
\label{eq:1t}
\end{eqnarray}
The Einstein equations are (\ref{eq:1r}) and (\ref{eq:1s}), whereas
(\ref{eq:1t}) is the YM equation. Our EDYM system is given by the five
ODEs (\ref{eq:1d1}), (\ref{eq:1d2}), (\ref{eq:1r})--(\ref{eq:1t}), 
together with the normalization condition~(\ref{eq:1nn}).

We are interested here in bound states of the Dirac particles. Thus we want to
find particle-like solutions of our EDYM system, i.e.\ solutions which are
smooth and tend to the vacuum solution as $r \rightarrow
\infty$. According to the explicit formulas (\ref{eq:TT0})--(\ref{eq:TT1}),
the energy-momentum
tensor of the YM field is regular at $r=0$ only when $|w(0)|=1$ and
$w'(0)=0$. Using the remaining gauge freedom, we can assume that
$w(0)=1$, and thus
\begin{equation}
w(r) \;=\; 1 \:-\: \frac{\lambda}{2} \:r^2 \:+\: {\cal{O}}(r^3)
\label{eq:22a}
\end{equation}
with a real parameter $\lambda$.
Using this result, a local Taylor expansion of the Einstein and Dirac
equations around $r=0$ yields (just as in~\cite{FSY1}) that
\begin{eqnarray}
\alpha(r) &=& \alpha_1 \:r \:+\: {\cal{O}}(r^3) \;\;\;,\spc
\beta(r) \;=\; \frac{1}{2} \:(\omega T_0 - m) \:\alpha_1 \:r^2
\:+\: {\cal{O}}(r^3) \label{eq:23} \\
A(r) &=& 1 \:+\: {\cal{O}}(r^2) \;\;\;\;\;\;\;\:,\spc
T(r) \;=\; T_0 \:+\: {\cal{O}}(r^2) \label{eq:24}
\end{eqnarray}
with two parameters $\alpha_1$ and $T_0>0$. Using linearity of the
Dirac equation, we can always assume that $\alpha_1>0$.
Furthermore, we demand that our solution has finite ADM mass,
\begin{equation}
\rho := \lim_{r \rightarrow \infty} \frac{r}{2 \kappa} (1-A(r)) \;<\; 
\infty \;\;\;, \label{eq:25xx}
\end{equation}
and goes asymptotically to the vacuum solution,
\begin{equation}
\lim_{r \rightarrow \infty} T(r) \;=\; 1 \;\;\;,\;\;\;\;\;
\lim_{r \rightarrow \infty} (w(r), w'(r)) \;=\; (\pm 1,0) \;\;\;,\;\;\;\;\;
\lim_{r \rightarrow \infty} (\alpha(r), \beta(r)) \;=\; (0,0) \;\;\;.
\label{eq:25}
\end{equation}

\section{The Reciprocal Coupling Limit}
\setcounter{equation}{0} \label{sec3}
Under all realistic conditions, the coupling of Dirac particles to
the YM field (describing the weak or strong interactions) is much stronger than
the coupling to the gravitational field. Thus we are particularly interested
in the case of weak gravitational coupling. In preparation, it is instructive
to briefly consider the case without gravitation. In this limit, the Dirac
equations read
\begin{eqnarray}
\alpha' &=& \frac{w}{r} \:\alpha \:-\: (m + \omega) \:\beta
\label{eq:26} \\
\beta' &=& (-m+\omega) \:\alpha \:-\: \frac{w}{r} \:\beta \;\;\;.
\label{eq:27}
\end{eqnarray}
For large $r$, these equations go over to a linear system of ODEs with 
constant coefficients, and the sign of $m-\omega$ determines whether the
solutions of these equations behave oscillatory or exponentially.
The normalization condition (\ref{eq:1nn}) excludes the oscillatory case
(as in~\cite[Section~5]{FSY3}) and thus $m-\omega \geq 0$ .
In the case $m-\omega=0$, the $\beta$-equation is
independent of $\alpha$, and the boundary conditions (\ref{eq:23}) imply
that $\beta \equiv 0$. As a consequence, (\ref{eq:1t}) reduces to the
homogeneous YM equation
\begin{equation}
r^2 \: w'' \;=\; -(1-w^2)\: w \;\;\;. \label{eq:2b}
\end{equation}
It is well-known~\cite{Y} that the only solution to this equation
satisfying the boundary conditions (\ref{eq:22a}),(\ref{eq:25}) is the
trivial solution $w \equiv 1$. But then the $\alpha$-equation simplifies to
\[ \alpha' \;=\; \frac{1}{r} \:\alpha \;\;\;, \]
whose solution $\alpha = \alpha_1 \:r$ violates the normalization condition
(\ref{eq:1nn}). In the case $m-\omega>0$, on the other hand, the local
Taylor expansion (\ref{eq:23}) yields that the $(\alpha, \beta)$-curve
lies for small $r$ in the fourth quadrant, i.e.\ $\beta(r)<0<\alpha(r)$ for
small $r$. Using the Dirac equations (\ref{eq:26}),(\ref{eq:27}), one
sees that the fourth quadrant is an invariant region, and thus
$\beta(r)<0<\alpha(r)$ for all $r$. But in the fourth quadrant, both
$\alpha(r)$ and $-\beta(r)$ are increasing for large $r$ (as one sees in
(\ref{eq:26}),(\ref{eq:27}) taking into account that $w/r \rightarrow 0$
for $r \rightarrow \infty$), and thus the normalization condition
(\ref{eq:1nn}) will again be violated.

These considerations show that the gravitational field is essential for
the formation of bound states. Nevertheless, for arbitrarily weak
gravitational coupling, we can hope to find bound states. It is even
conceivable that these bound state solutions might have a well-defined limit
when the gravitational coupling tends to zero, if we let the YM coupling go to
infinity at the same time. Our idea is that this limiting case might yield
a system of equations which is simpler than the full EDYM system, and can
thus serve as a physically interesting starting point for the analysis of the
coupled interactions described by the EDYM equations.
Expressed in dimensionless quantities, we shall thus
consider the limits
\begin{equation}
m^2 \kappa \;\rightarrow\; 0 \spc {\mbox{and}} \spc
e^2 \;\rightarrow\; \infty \;\;\;.
\label{eq:2A}
\end{equation}
Let us determine how the quantities of our EDYM system should behave in this
limit. Since we are considering weak gravitational coupling, it is clear that
the metric will be close to the Minkowski metric, i.e.\ $A \approx 1$ and
$T \approx 1$. Furthermore, the YM potential $w$ should have a finite 
limit. Similar to our flat space consideration at
the beginning of this section, one sees that the normalization condition
(\ref{eq:1nn}) can be satisfied only if the function $m-\omega T(r)$ changes
sign, and thus $\omega \approx m$ (but both $m$ and
$\omega$ may go to zero or infinity in the limit (\ref{eq:2A})). Putting this
information together, we conclude that the Dirac equations (\ref{eq:1d1})
and (\ref{eq:1d2}) have a meaningful limit only under the assumptions that
$\alpha$ converges and that
\begin{equation}
m \:\beta(r) \;\rightarrow\; \hat{\beta}(r) \;\;\;,\spc
m^2 \:(T(r)-1) \;\rightarrow\; \varphi \;\;\;,\spc
m\:(\omega-m) \;\rightarrow\; E
\label{eq:2B}
\end{equation}
with two real functions $\hat{\beta}$, $\varphi$ and a real parameter $E$.
Multiplying (\ref{eq:1d2}) with $m$ and taking the limits (\ref{eq:2B}) as
well as $A, T \rightarrow 1$, the Dirac equations become
\begin{eqnarray}
\alpha' &=& \frac{w}{r} \:\alpha \:-\: 2 \:\hat{\beta}
\label{eq:2l1} \\
\hat{\beta}' &=& (E+\varphi) \:\alpha \:-\: \frac{w}{r}
\:\hat{\beta} \;\;\;.
\label{eq:2l2}
\end{eqnarray}
We next consider the YM equation~(\ref{eq:1t}). The last term in
(\ref{eq:1t}) drops out in the limit of weak gravitational coupling
(\ref{eq:2A}). The second summand converges only under the assumption that
\begin{equation}
\frac{e^2}{m} \;\rightarrow\; q \label{eq:6a}
\end{equation}
with $q$ a real parameter, playing the role of an ``effective''
coupling constant. Together with (\ref{eq:2A}), this implies that
$m \rightarrow \infty$. The YM equations thus have the limit
\begin{equation}
r^2\:w'' \;=\; -(1-w^2)\:w \:+\: q \:r\: \alpha \hat{\beta}
\;\;\;. \label{eq:2l3}
\end{equation}
In order to get a well-defined and non-trivial limit of the
Einstein equations (\ref{eq:1r}),(\ref{eq:1s}), we need to
assume that the parameter $m^3 \kappa$ has a finite, non-zero limit.
Since this parameter has the
dimension of inverse length, we can arrange by a scaling of our coordinates
that
\begin{equation}
m^3 \kappa \;\rightarrow\; 1 \;\;\;. \label{eq:7a}
\end{equation}
We differentiate the $T$-equation
(\ref{eq:1s}) with respect to $r$ and substitute (\ref{eq:1r}).
Taking the limits (\ref{eq:2B}) and (\ref{eq:7a}), a straightforward
calculation yields the equation
\begin{equation}
r^2 \:\Delta \varphi \;=\; -\alpha^2 \;\;\;,
\label{eq:2l4}
\end{equation}
where $\Delta=r^{-2} \partial_r (r^2 \partial_r)$ is the radial Laplacian
in Euclidean $\R^3$. Indeed, this equation can be regarded as Newton's
equation with the Newtonian potential $\varphi$. Thus our limiting
case (\ref{eq:2l4}) for the gravitational field corresponds to taking the
Newtonian limit. Finally, the normalization condition (\ref{eq:1nn})
reduces to
\begin{equation}
\int_0^\infty \alpha(r)^2 \:dr \;=\; 1 \;\;\;.
\label{eq:2l5}
\end{equation}
The boundary conditions (\ref{eq:22a})--(\ref{eq:25}) are transformed into
\begin{eqnarray}
w(r) &=& 1 \:-\: \frac{\lambda}{2} \:r^2 \:+\: {\cal{O}}(r^3) \;\;\;,\spc
\lim_{r \rightarrow \infty} w(r) \;=\; \pm 1 \label{eq:9b} \\
\alpha(r) &=& \alpha_1\: r \:+\: {\cal{O}}(r^3) \spc\:,\spc
\hat{\beta}(r) \;=\; {\cal{O}}(r^3) \label{eq:9c} \\
\varphi(r) &=& \varphi_0 \:+\: {\cal{O}}(r^2) \spc\;\;\;,\spc
\lim_{r \rightarrow \infty} \varphi(r) \;<\; \infty \label{eq:9d}
\end{eqnarray}
with the three parameters $\lambda$, $\alpha_1$, and $\varphi_0$.
We point out that the limiting system contains only one coupling constant $q$.
According to (\ref{eq:6a}) and (\ref{eq:7a}), $q$ is in dimensionless
form given by
\begin{equation}
e^2 \:m^2 \kappa \;\rightarrow\; q \;\;\;. \label{eq:13a}
\end{equation}
Hence in dimensionless quantities, our limit (\ref{eq:2A})
describes the situation where the 
gravitational coupling goes to zero, while the YM coupling constant 
goes to infinity like $e^2 \sim (m^2 \kappa)^{-1}$. Therefore, we 
call our limiting case the {\em{reciprocal coupling limit}}. The 
reciprocal coupling system is given by the equations (\ref{eq:2l1}),
(\ref{eq:2l2}), (\ref{eq:2l3}), and (\ref{eq:2l4}) together with the
normalization condition (\ref{eq:2l5}) and the boundary conditions
(\ref{eq:9b})--(\ref{eq:9d}). According to
(\ref{eq:2B}), the parameter $E$ coincides up to a scaling factor with
$\omega-m$, and thus has the interpretation as the (properly scaled)
energy of the Dirac particle. As in Newtonian mechanics, the potential
$\varphi$ is determined only up to a constant $\mu \in \R$; namely, the
reciprocal limit equations are invariant under the transformation
\begin{equation}
\varphi \;\rightarrow\; \varphi + \mu \;\;\;,\spc
E \;\rightarrow\; E - \mu \;\;\;. \label{eq:9a}
\end{equation}

Let us consider how the ADM mass behaves in the reciprocal coupling 
limit. First of all, we can write the quotient $\rho/m$ as
\[ \frac{\rho}{m} \;=\; \lim_{r \rightarrow \infty} \frac{r}{2 \kappa m}
\:(1-A(r)) \;=\; \frac{1}{m} \int_0^\infty \left(\frac{r}{2 \kappa}\:
(1-A(r)) \right)' \:dr\;\;\;. \]
After substituting the $A$-equation (\ref{eq:1r}), we can take the limits
(\ref{eq:2A}) and (\ref{eq:2B}) and obtain that
\begin{equation}
\frac{\rho}{m} \;\rightarrow\; \int_0^\infty \alpha(r)^2 \:dr
\;\stackrel{(\ref{eq:2l5})}{=}\; 1 \;\;\;. \label{eq:r1}
\end{equation}
Thus the ADM mass coincides with the rest mass of the Dirac particle; this
shows that the total binding energy $B:=\rho-m$ goes to zero in our limit.
Indeed, the behavior of the total binding energy can be described in more
detail as follows. For a solution of the full EDYM system, we can write the
binding energy using the normalization condition (\ref{eq:1nn}) as
\[ B \;=\; \int_0^\infty \left( \left(\frac{r}{2 \kappa} \:(1-A)\right)'
- m \:(\alpha^2 + \beta^2) \:\frac{T}{\sqrt{A}} \right) dr \;\;\;. \]
We again substitute the $A$-equation (\ref{eq:1r}) and obtain
\begin{equation}
B \;=\; \int_0^\infty \left(\frac{1}{2e^2} \:\frac{(1-w^2)^2}{r^2} \:+\:
\frac{1}{e^2} \: A w'^2 \:+\: (w T \sqrt{A} - m) \:\frac{T}{\sqrt{A}} \:
(\alpha^2+\beta^2) \right) dr \;\;\;. \label{eq:r2}
\end{equation}
According to (\ref{eq:13a}), it is obvious that the first two summands
in (\ref{eq:r2}) have a finite limit after dividing by $m^2 \kappa$. In order
to treat the last summand, we first multiply the $T$-equation (\ref{eq:1s})
with $m^2$ and take the limits (\ref{eq:2A}), (\ref{eq:2B}),
(\ref{eq:7a}),
\[ m^2 \:(A-1) \;\rightarrow\; 2r \:\varphi' \;\;\;. \]
Using again (\ref{eq:7a}), (\ref{eq:2B}), and $\omega \rightarrow m$,
$T \rightarrow 1$, we obtain that
\begin{eqnarray*}
\frac{1}{m^2 \kappa} \:(\omega T \sqrt{A} - m) &=& \frac{1}{m^3 \kappa} \:
m \omega \:(\sqrt{A}-1) \:T \:+\: \frac{1}{m^3 \kappa} \:m(\omega T - m) \\
&\rightarrow& r \varphi' + (\varphi + E) \;\;\; .
\end{eqnarray*}
From this we conclude that the binding energy (\ref{eq:r2}) divided by
$m^2 \kappa$ has a meaningful limit; more precisely
\begin{equation}
\hat{B} \;:=\; \frac{B}{m^2 \kappa} \;\rightarrow\;
\int_0^\infty \left(\frac{1}{2 q} \:\frac{(1-w^2)^2}{r^2} \:+\:
\frac{1}{q} \: w'^2 \:+\: \alpha^2 \:(E+\varphi+r \varphi')
\right) dr \;\;\;. \label{eq:r3}
\end{equation}\\

We now describe our method for constructing numerical solutions of our
reciprocal limit system. Since it is difficult to take into account the
integral condition (\ref{eq:2l5}) in the numerics, we discard this
condition for the construction of the solution; it
will be taken care of later via a rescaling technique (see (\ref{eq:G1}),
(\ref{eq:G2})). This rescaling method requires only that the normalization
integral be finite,
\begin{equation}
0 \;<\; \lambda^2 := \int_0^\infty \alpha(r)^2 \:dr \;<\; \infty \;\;\;.
\label{eq:E}
\end{equation}
According to (\ref{eq:2l1}), (\ref{eq:2l2}) and (\ref{eq:9b}), (\ref{eq:9d}),
the behavior of the Dirac spinors at infinity is either oscillatory or
exponential. As a consequence, the normalization integral in (\ref{eq:E})
will be finite only if $\alpha(r)$ tends to zero for $r \rightarrow \infty$.
Furthermore, we can use the transformation (\ref{eq:9a}) to set
$\varphi(0)=0$. Hence in the first construction
step, we want to find solutions of the modified system
\begin{eqnarray}
\alpha' &=& \frac{w}{r} \:\alpha \:-\: 2 \:\hat{\beta}
\label{eq:F1} \\
\hat{\beta}' &=& (E+\varphi) \:\alpha \:-\: \frac{w}{r}
\:\hat{\beta} \label{eq:F2} \\
r^2\:w'' &=& -(1-w^2)\:w \:+\: q \:r\: \alpha \hat{\beta}
\label{eq:F3} \\
r^2 \:\Delta \varphi(r) &=& -\alpha^2 \label{eq:F4}
\end{eqnarray}
with the following conditions at the origin,
\begin{eqnarray}
w(r) &=& 1 - \lambda r^2 + {\cal{O}}(r^3) \;\;\;,\spc
\alpha(r) \;=\; \alpha_1\: r + {\cal{O}}(r^2) \label{eq:A} \\
\varphi(r) &=& {\cal{O}}(r^2) \spc\;\;\;\;\;\;\;\;\;,\spc
\beta(r) \;=\; {\cal{O}}(r^3) \;\;\;, \label{eq:B}
\end{eqnarray}
together with the conditions at infinity
\begin{eqnarray}
\lim_{r \rightarrow \infty} w(r) &=& \pm 1
\;\;\;,\spc \lim_{r \rightarrow \infty} \alpha(r) \;=\; 0 \label{eq:C} \\
|\varphi_\infty| := |\lim_{r \rightarrow \infty} \varphi(r)| &<& \infty
\;\;\;. \label{eq:D}
\end{eqnarray}
For any given value of the coupling constant $q$, we thus have two free
parameters $\lambda$ and $\alpha_1$ to characterize the solutions near the
origin $r=0$. Each solution has a unique extension to larger values of $r$.
Asymptotically for $r \rightarrow \infty$, we must satisfy the two conditions
(\ref{eq:C}). Thus we have as many free parameters as asymptotic conditions,
and we therefore expect for fixed $q$ a discrete set of solutions satisfying
(\ref{eq:A}), (\ref{eq:B}), and (\ref{eq:C}).
For each solution, we must then verify that the conditions (\ref{eq:E}) and
(\ref{eq:D}) are also satisfied.

For the construction of numerical solutions, we enhanced the technique
used in \cite{FSY1, FSY2} to a two-parameter shooting method. Since
two-parameter problems are considerably more difficult than one-parameter
problems, we describe the method in some detail.
For clarity, we first consider the simplified situation where $\alpha(r)$ and
$w(r)$ have prescribed boundary values for a given finite $r=r_1$.
In this case, one can apply the standard multi-parameter shooting method as
e.g.\ described in \cite{S}. More precisely, one can for given initial data
compute $\alpha(r_1)$ and $(w(r_1), w'(r_1))$ numerically, compare with
the prescribed boundary conditions, and iteratively adjust the two
free parameters $\lambda$ and $\alpha_1$
until the boundary conditions are satisfied to sufficient accuracy.
In our case, the situation is more difficult because we have boundary values
not for finite $r=r_1$, but for $r=\infty$. In order to deal with this problem,
we first choose a finite $r_1$. Using an ansatz for the asymptotic form of
the solution $(\alpha, \hat{\beta}, w, \varphi)$ at infinity, we
approximately compute $\alpha(r_1)$ and $(w(r_1), w'(r_1))$ and derive
conditions between these functions.
Taking these conditions as the boundary conditions at
$r=r_1$, we can apply the two-parameter shooting method on the finite
interval $(0, r_1]$ as described above. The so-obtained solution on
$(0,r_1]$ gives, in combination with the asymptotic formulas on
$(r_1, \infty)$, an approximate solution for all $r>0$. Since our
asymptotic description becomes precise only in the limit $r_1 \rightarrow
\infty$, we must, in order to attain the desired accuracy, choose $r_1$
sufficiently large. In order to ensure that $r_1$ is appropriately increased
during the computation, we modified the two-parameter shooting method in such
a way that both the initial data and $r_1$ are adjusted in each iteration
step. The iteration is stopped
when the numerics has stabilized and the accuracy no longer improves.
This modified shooting method was implemented in the Mathematica programming
language using the standard ODE solver with a working precision of 16 digits.
The initial data is adjusted in the iteration with a secant method,
and the step size for incrementing $r_1$ is determined from the relative
error of the numerical solution at the upper boundary $r_1$.
After the iteration has been stopped and a numerical solution has been found,
our program slightly changes the initial data and searches for a nearby
solution. In this way, we can automatically trace a one-parameter family
of solutions.
Finally, we explain our method for describing the asymptotic behavior of
the solutions at infinity. According to the asymptotics of the solutions of
the ED and EYM equations \cite{FSY1, BM}, we can expect that the spinors
$\alpha$ and $\hat{\beta}$ will decay exponentially fast at infinity, whereas
the potentials $\varphi(r)$ and $w(r)$ for $r \rightarrow \infty$ will
behave like rational functions. Therefore it is a reasonable asymptotic
approximation to set $\alpha$ and $\hat{\beta}$ to zero for $r>r_1$.
In this approximation, the potential $\varphi$ is harmonic according to
(\ref{eq:F4}). The YM equation (\ref{eq:F3}), on the
other hand, reduces to the vacuum YM equation (\ref{eq:2b}).
In the new variable $u=\log r$, this equation becomes autonomous;
namely \cite{Y}
\begin{equation}
\partial_u^2 w - \partial_u w \;=\; -(1-w^2) \:w \;\;\;.
\label{eq:x}
\end{equation}
This autonomous equation allows us to derive boundary conditions for $w$
as follows. We set
\begin{equation}
x = w \spc {\mbox{and}} \spc y = \partial_u w \;\;\;. \label{eq:y}
\end{equation}
Then the YM orbits in the $(x,y)$ plane are described by the following
differential equation,
\begin{equation}
y'(x) \;=\; \frac{\partial_u^2 w}{\partial_u w} \;\stackrel{(\ref{eq:x})}{=}\;
1 - \frac{(1-x^2)\:x}{y} \;\;\;. \label{eq:z}
\end{equation}
According to the boundary conditions (\ref{eq:C}) and the differential equation
(\ref{eq:x}), the variables $x$ and $y$ must behave in the limit $r \rightarrow
\infty$ like either $x \rightarrow 1$, $y \searrow 0$ or
$x\rightarrow -1$, $y \nearrow 0$. In both of these cases, there is a 
unique YM orbit $y(x)$, which can be easily calculated numerically by
integrating (\ref{eq:z}). 
By transforming (\ref{eq:y}) back to the variable $r$, we obtain
the following mixed boundary conditions for $w(r)$ at $r=r_1$,
\begin{equation}
w'(r_1) \;=\; \frac{1}{r_1} \:y(w(r_1)) \;\;\;. \label{eq:bc}
\end{equation}

We next describe our rescaling method needed to arrange the
normalization condition (\ref{eq:2l5}). Suppose that we have a solution
of the modified system (\ref{eq:F1})--(\ref{eq:D}) with finite normalization
integral, (\ref{eq:E}). A direct calculation shows that the transformed
functions
\begin{eqnarray}
\tilde{\alpha}(r) &=& \lambda^{-2} \:\alpha(\lambda^{-2} r)
\spc\;\;\;\;\;\;\:\:\!,\spc
\tilde{\hat{\beta}}(r) \;=\; \lambda^{-4} \:\beta(\lambda^{-2} r)
\label{eq:G1} \\
\tilde{\varphi}(r) &=& \lambda^{-4} \left(
\varphi(\lambda^{-2} r) - \varphi_\infty \right) \;\;\;,\spc
\tilde{w}(r) \;=\; w(\lambda^{-2} r) \label{eq:G2}
\end{eqnarray}
solve our original reciprocal limit system (\ref{eq:2l1}),
(\ref{eq:2l2}), (\ref{eq:2l3}), (\ref{eq:2l4}), and (\ref{eq:2l5})
with boundary conditions (\ref{eq:9b})--(\ref{eq:9d}), if one
replaces the energy $E$ and coupling constant $q$ by
\begin{equation}
\tilde{E} \;=\; \lambda^{-4} \left(E + \varphi_\infty \right)
\spc{\mbox{and}}\spc \tilde{q} \;=\; \lambda^4 \:q \;\;\;. \label{eq:G3}
\end{equation}
We point out that only the rescaled solutions (\ref{eq:G1}),(\ref{eq:G2})
and rescaled parameters (\ref{eq:G3}) have a physical meaning.
Therefore, we will in what follows consider only the rescaled tilde
solutions; for ease in notation, the tilde will be omitted.\\

In the remainder of this section, we describe our numerical solutions
of the reciprocal limit equations. Just as in the case for
the ED and EYM equations \cite{FSY1, BM}, there are solutions for different
rotation numbers of the spinors in the $(\alpha, \beta)$-plane and for the
YM potential in the $(w, w')$-plane. For simplicity, we restricted
attention to solutions with rotation number zero for the spinors (as for
the ground state solutions in \cite{FSY1}). For the YM potential, we 
consider only the cases where the $(w, w')$-curve either makes a 
half rotation joining the points $(1,0)$ and $(-1,0)$, or makes a 
full rotation, ending at its starting point $(1,0)$.
A typical example for a solution of each
type is shown in Figures~\ref{limit1} and~\ref{limit2}. 
Because of the similarity of the YM potential to the BM ground state 
and the BM first excited state, we refer to these two types in what follows
as the ground states and the first excited states, respectively.
Notice that the curves in the $(w, w')$-plane are not
plotted all the way to their rest points at $(1,0)$ or $(0,-1)$, respectively.
The reason is that we plot only the numerical solution on the interval
$(0,r_1]$. One sees that the spinors have become
practically zero for $r=r_1$, and it is thus an admissible approximation
to smoothly join the $(w,w')$-curve with a vacuum YM solution by using
the boundary conditions (\ref{eq:bc}).
We first discuss the ground state solutions. In Figure~\ref{limit3}, 
the main characteristics of the solutions are plotted versus
the coupling constant $q$. As explained above, $E$
has the interpretation of the (appropriately scaled) energy of the
Dirac particle. Since $E$ is negative, the Dirac particle has gained energy
by forming the bound state. The parameter $\hat{B}$, (\ref{eq:r3}), gives
the total binding energy, i.e.\ the amount of energy which is set free when
the binding is broken up. Since $\hat{B}$ is negative, we can expect that
solutions of the full EDYM system, which are close to the
solutions of the reciprocal coupling equations, should be stable.
Finally, $r_w$ and $r_\alpha$ are the characteristic length scales of the
solutions; more precisely, $r_w$ is the radius where $w$ changes sign,
and $r_\alpha$ is the radius where $\alpha$ has its maximum,
\begin{equation}
w(r_w) \;=\; 0 \spc{\mbox{and}}\spc \alpha'(r_\alpha) \;=\; 0 
\;\;\;. \label{eq:rw1}
\end{equation}
The characteristic radii are interesting because they give information about
the ``size'' of the solutions as functions of $r$; i.e.\ they tell whether
the fields are spread out in space, or whether they are localized close to
the origin. It is also worth considering both radii because $r_w$ and
$r_\alpha$ can behave quite differently (cf.\ Figure~\ref{limit3}).

The plots in Figure~\ref{limit3} have a turning point at $q \approx 8.49$.
Similar to the situation described for the spiral in~\cite{FSY1}, this is a
bifurcation point which comes about as a consequence of our rescalings. One
branch of solutions can be extended up to $q \approx 11.6$. For solutions
close to this end point, the potential $w$ leaves the interval $[-1,1]$ as
shown in Figure~\ref{limit5}.
Since $r_w$ and $r_\alpha$ both go to zero in this
limit, the spinors and YM field are both confined to a smaller and smaller
neighborhood of the origin. At the same time, the energy of the Dirac
particle and the binding energy become infinite.
The other branch of solutions ends near $q=8.95$. For solutions near this
end point, the $(w, w')$-curve comes very close to the origin before
running into the rest point at $(-1,0)$, see Figure~\ref{limit6}.
This makes the numerics rather
delicate, and we therefore have not yet analyzed this regime in much detail.
It is interesting that $r_\alpha$ is bounded near this end point, whereas
$r_w$ seems to become infinite. This shows that, while the Dirac particle
stays in a bounded region of space, the YM field becomes more and more spread
out.

For the first excited state, the energy spectrum and characteristic radii are
shown in Figure~\ref{limit7}.
Since in general $w$ never equals zero for
the first excited state, we define $r_w$ via the minimum of $w$, i.e.
\begin{equation}
w'(r_w) \;=\; 0 \spc{\mbox{and}}\spc \alpha'(r_\alpha) \;=\; 0 
\;\;\;. \label{eq:rw2}
\end{equation}
In contrast to the ground state, the solutions can now be
extended up to $q=0$. In this regime, the YM potential stays close to
$w=1$; see Figure~\ref{limit8}.
The solutions have a bifurcation point
at $q \approx 9.866$. The branch coming out at the bifurcation point for
larger values of $E$ is difficult to study numerically because the
$(w,w')$-curve comes close to the origin, see Figure~\ref{limit9}.

It is interesting that for the ground state solutions in Figure~\ref{limit3},
the parameter $q$ stays bounded away from zero, whereas the plots for 
the first excited state in Figure~\ref{limit7} could be extended up 
to $q=0$. Let us consider how this can be understood directly from the 
equations. The parameter $q$ enters only the YM equation 
(\ref{eq:2l3}). In the limit $q \rightarrow 0$, this equation goes 
over to the vacuum YM equation, which has only the trivial solution 
$w \equiv 1$. Hence if we assume that the spinors have a finite limit for $q 
\rightarrow 0$, then $w(r)$ must go uniformly in $r$ to one.
This shows that the solutions can be regular for $q \rightarrow 0$
only if $w$ satisfies the boundary condition $\lim_{r \rightarrow \infty}
w(r)=+1$. In particular, our ground state solutions cannot be regular in this 
limit. We next consider the limit $q \rightarrow 0$ 
for the first excited state in more detail. Since $w$ converges 
uniformly in $r$ to one, the reciprocal limit equations (\ref{eq:2l1}),
(\ref{eq:2l2}), 
and (\ref{eq:2l4}) go over to the Dirac-Newton equations
\begin{eqnarray}
\alpha' &=& \frac{1}{r} \:\alpha \:-\: 2 \:\hat{\beta}
\label{eq:DN1} \\
\hat{\beta}' &=& (E+\varphi) \:\alpha \:-\: \frac{1}{r}
\:\hat{\beta} \\
r^2 \:\Delta \varphi(r) &=& -\alpha^2 \;\;\;.
    \label{eq:DN3}
\end{eqnarray}
These equations are obtained by taking the nonrelativistic limit of 
the ED equations \cite{FSY1}, and according to the results obtained in 
that paper, the equations (\ref{eq:DN1})--(\ref{eq:DN3}) together with 
the normalization integral (\ref{eq:2l5}) have a countable number of 
solutions, characterized by the rotation number of the spinors (called 
the ground state, the first excited state, etc.). We thus expect that 
the functions $(\alpha, \hat{\beta}, \varphi)$ corresponding to 
solutions of the reciprocal limit equations should for $q \rightarrow 0$, go 
over to a solution of the Dirac-Newton equations.
The behavior of the YM potential $w$ can now be analyzed in more 
detail by taking the solution $(\alpha, \hat{\beta}, \varphi)$ of the 
Dirac-Newton equations as a given inhomogeneity in the YM equation 
(\ref{eq:2l3}) and performing a perturbation calculation for small $q$. 
More precisely, the ansatz $w(r)=1+q \:u(r)$ to first order in 
$q$, leads to the linear equation
\[ r^2 \:u'' \;=\; 2 u \:+\: r\: \alpha \hat{\beta} \;\;\;, \]
which can be solved by integration. Fixing the integration constants 
with our boundary conditions $u(0)=u(\infty)=0$ and $u'(0)=0$, we 
obtain the unique solution
\begin{eqnarray}
u(r) \;=\; r^2 \int_0^r \frac{ds}{s^4} \int_0^s t \:\alpha(t) 
\:\hat{\beta}(t) \:dt \:-\: \frac{r^2}{3} \int_0^\infty 
\frac{1}{t^2} \:\alpha(t)\: \hat{\beta}(t)\:dt \;\;\;.
\label{eq:DN4}
\end{eqnarray}
This consideration shows that for $q \rightarrow 0$, the rotation 
number of $w$ is uniquely determined by the rotation number of the 
spinors. Furthermore, one sees that in the limit $q \rightarrow 0$, the
Dirac wave function is determined by the Dirac-Newton equations
(\ref{eq:DN1})--(\ref{eq:DN3}). Thus only the gravitational attraction is
responsible for the formation of the bound state, whereas the YM field
has no influence on the spinors.

\section{Solutions of the EDYM Equations}
\setcounter{equation}{0}
In this section, we shall construct numerical solutions of the full EDYM 
equations and discuss their properties. Our method is to first find 
special solutions which are small perturbations of either the BM 
solutions \cite{BM} or solutions to the reciprocal limit equations of 
the previous section. We then trace these solutions while gradually 
changing the coupling constants. This yields one-parameter families of 
solutions which can be extended even to regions in parameter space 
where the solutions are far from all of the known limiting cases.

In order to simplify the connection between the EDYM equations and the 
reciprocal limit equations of Section~\ref{sec3}, it is useful to introduce a 
parameter $\varepsilon>0$ in such a way that the reciprocal limit equations 
are obtained when $\varepsilon \rightarrow 0$. To this end, we 
parametrize the EDYM equations in terms of the new variables 
$(\varepsilon, q, E)$ as follows,
\begin{eqnarray*}
\kappa & = & (\varepsilon q)^{\frac{3}{2}} \;\;\;,\spc
e^2 \;=\; \sqrt{\frac{q}{\varepsilon}} \\
m & = & \frac{1}{\sqrt{\varepsilon q}} \;\;\;\:\:,\spc
\omega \;=\; \frac{1}{\sqrt{\varepsilon q}} \:+\: \sqrt{\varepsilon 
q} \:E \;\;\;.
\end{eqnarray*}
Since the EDYM equations involve three dimensionless parameters 
(namely $m^2 \kappa$, $\omega/m$, and $e^2$), introducing 
$(\varepsilon, q, E)$ is merely a transformation to new independent 
parameters, prescribing at the same time the gravitational constant 
(this means that we give up the freedom to rescale $r$ by fixing our 
length scale). In the limit $\varepsilon \rightarrow 0$, both $q$ and $E$ 
go over to the corresponding parameters of the reciprocal limit system (see 
(\ref{eq:6a}) and (\ref{eq:2B})). Also, it is easy to check that the 
limits (\ref{eq:2A}), (\ref{eq:7a}), and (\ref{eq:13a}) are satisfied 
if we let $\varepsilon \rightarrow 0$ and keep $(q, E)$ fixed. The 
parameters $\varepsilon$ and $q$ can be written in dimensionless 
form as
\begin{equation}
\varepsilon \;=\; \frac{m^2 \kappa}{e^2} \;\;\;,\spc
q \;=\; m^2 \kappa\: e^2 \;\;\;.
    \label{eq:cp1}
\end{equation}
Thus $\varepsilon$ describes the relative strength of gravity 
versus the YM interaction, whereas $q$ is the product of the 
gravitational and YM coupling constants. Up to a scale factor,
$E=\omega-m$. Since $\omega$ is the relativistic energy and
$m$ the rest mass of the Dirac particle, $E$ can, exactly as in
the previous section, be interpreted as the
energy of the Dirac particle. Finally, we 
also describe the binding energy by a parameter which corresponds to
our notation for the reciprocal limit system (\ref{eq:r3}) and set
\[ \hat{B} \;=\; \frac{\rho - m}{\sqrt{\varepsilon q}} \;\;\;. \]

For the construction of numerical solutions, we use a two-parameter 
shooting algorithm combined with a rescaling method. Since this 
technique is quite similar to that described for the reciprocal limit 
equations in the previous section, we shall merely outline our procedure. In 
the first step of the construction, we consider the EDYM equations
(\ref{eq:1d1}), (\ref{eq:1d2}) and (\ref{eq:1r})--(\ref{eq:1t}) with 
the side conditions
\begin{eqnarray}
0 &<& \lambda^2 := \int_0^\infty (\alpha^2 + \beta^2) 
\:\frac{T}{\sqrt{A}} \:dr \;<\; \infty \label{eq:cn1} \\
0 &<& \tau := \lim_{r \rightarrow \infty} T(r) \;<\; \infty 
\label{eq:cn2} \\
\lim_{r \rightarrow \infty} w(r) &=& \pm 1 \;\;\;, \label{eq:cn3}
\end{eqnarray}
together with the following expansions near $r=0$,
\begin{eqnarray*}
\alpha(r) &=& \alpha_1 \:r \:+\: {\cal{O}}(r^3) \;\;\;,\spc
\beta(r) \;=\; {\cal{O}}(r^2) \\
A(r) &=& 1 \:+\: {\cal{O}}(r^2) \;\;\;\;\;\;\;,\spc
T(r) \;=\; 1 \:+\: {\cal{O}}(r^2) \;\;\;.
\end{eqnarray*}
For fixed $\varepsilon$ and $q$, we thus have the two parameters 
$\alpha_1$ and $E$ to characterize a solution of this modified EDYM 
system near the origin $r=0$. On the other hand, we must satisfy two 
conditions at infinity; namely, $w$ must converge to $\pm 1$, 
(\ref{eq:cn3}), and the spinors must go asymptotically to zero in 
order for the normalization integral to be finite (\ref{eq:cn1}). 
Hence, we can apply a two-parameter shooting method as described in 
Section~\ref{sec3}. In order to have optimal boundary conditions at 
the upper end point $r=r_1$, we again match this with the solution of the
autonomous vacuum YM equation (see (\ref{eq:z}) and (\ref{eq:bc})).
The shooting method was again implemented in Mathematica, using an 
accuracy of 32 digits.
For each solution constructed in this way, we verify 
that (\ref{eq:cn2}) is satisfied and that the ADM mass is finite 
(\ref{eq:25xx}). Once we have found a solution of the modified 
equations, we rescale the solution according to
\begin{eqnarray}
\tilde{\alpha}(r) &=& \sqrt{\tau} \:\lambda^{-2} \:\alpha(\lambda^{-2} 
r) \;\;\;, \spc \tilde{\beta}(r) \;=\; \sqrt{\tau} \:\lambda^{-2} 
\:\beta(\lambda^{-2} r) \label{eq:res0} \\
\tilde{A}(r) &=& A(\lambda^{-2} r) \spc\;\;\;\;\;,\spc
\tilde{T}(r) \;=\; \tau^{-1} \:T(\lambda^{-2} r) \;\;\;, 
\end{eqnarray}
and transform the parameters $(m, \omega, \kappa, e^2)$ as follows,
\begin{eqnarray}
\tilde{m} &=& \lambda^{-2} \:m \;\;\;, \spc
\tilde{\omega} \;=\; \tau \:\lambda^{-2} \:\omega \\
\tilde{\kappa} &=& \lambda^6 \:\kappa \;\;\;\;\;\:,\spc
\tilde{e}^2 \;=\; \lambda^2 \:e^2 \;\;\;.
\end{eqnarray}
A straightforward calculation shows that the so-rescaled solution 
satisfies the EDYM equations (\ref{eq:1d1}), (\ref{eq:1d2}) and
(\ref{eq:1r})--(\ref{eq:1t}) together with the original side 
conditions (\ref{eq:1nn}) and (\ref{eq:23})--(\ref{eq:25}).
The parameters $(\varepsilon, q, E)$ transform under the rescalings as
\begin{equation}
\tilde{\varepsilon} \;=\; \varepsilon \;\;\;,\;\;\;\;\; \tilde{q} 
\;=\; \lambda^4 \:q \;\;\;,\;\;\;\;\; \tilde{E} \;=\; \lambda^{-4} 
\:(E + (\tau-1) \:m \omega) \;\;\;.
    \label{eq:res}
\end{equation}
In the limit $\varepsilon \rightarrow 0$, these transformations 
coincide with the rescalings of the reciprocal limit equations 
(\ref{eq:G1})--(\ref{eq:G3}). However, we remark that for the ED 
equations~\cite{FSY1} a much different rescaling is used. Namely, in 
order to get a better correspondence to the reciprocal limit equations, we 
here scale the gravitational constant $\kappa$, whereas in~\cite{FSY1}
$\kappa$ is fixed to be 1 throughout.
Clearly, only the rescaled solutions have physical significance. 
Therefore, in what follows we will consider only the rescaled 
solutions and again omit the tilde.

In a realistic physical situation, the gravitational coupling is very 
weak, whereas the YM coupling constant is of order one, i.e.
$m^2 \kappa \ll 1$ and $e^2 \sim 1$. Hence, 
according to (\ref{eq:cp1}), we will here only investigate the 
parameter range $\varepsilon \ll 1$, and we are particularly interested in 
the situation for small $q$. In the limit $\varepsilon \rightarrow 0$, 
there are known solutions of our EDYM system, namely the BM solutions 
(which more precisely are solutions in the limits $\kappa \rightarrow 0$ and 
$e^2/\kappa \rightarrow 1$), and the solutions of the reciprocal limit system 
constructed in Section~\ref{sec3}. We take these special solutions as 
the starting point for the numerics.
By varying the parameters $\varepsilon$ and $q$ 
and tracing the solutions with our shooting and rescaling methods, we 
obtain a two-parameter family of solutions. In order to reduce the 
computational workload, we did not step systematically through the 
two-parameter space, but always kept one parameter fixed while varying 
the other parameter. Since $\varepsilon$ remains unchanged under the 
rescaling (see (\ref{eq:res})), it is most convenient to construct 
one-parameter families of solutions for different, fixed values of 
$\varepsilon$.

\begin{figure}[thb]
	\centerline{\epsfbox{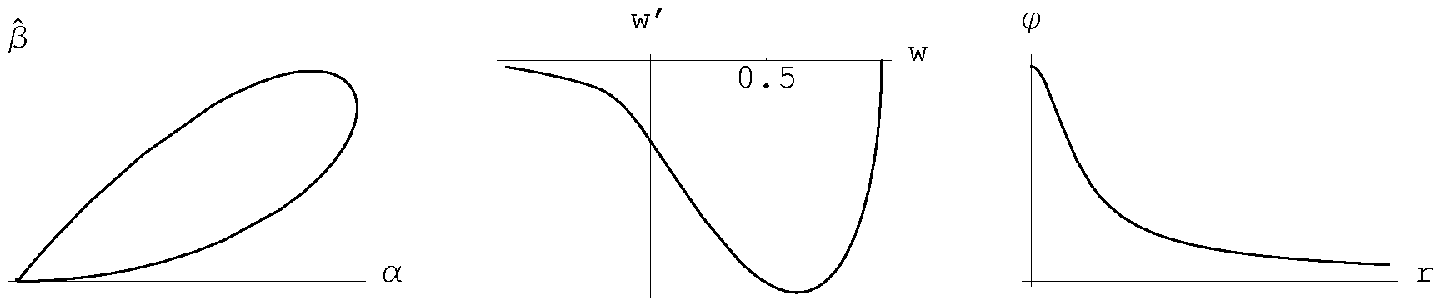}}
	\caption{Reciprocal coupling limit:
The ground state for $q=8.49811$, $E=0.377446$.}
	\label{limit1}
\centerline{} \centerline{}
	\centerline{\epsfbox{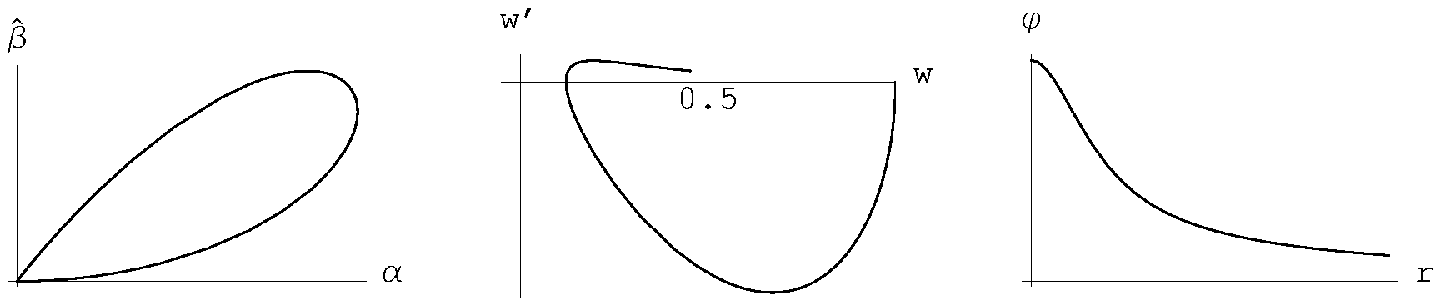}}
	\caption{Reciprocal coupling limit:
The first excited state for $q=6.96132$, $E=0.227762$.}
	\label{limit2}
\centerline{} \centerline{}
	\centerline{\epsfbox{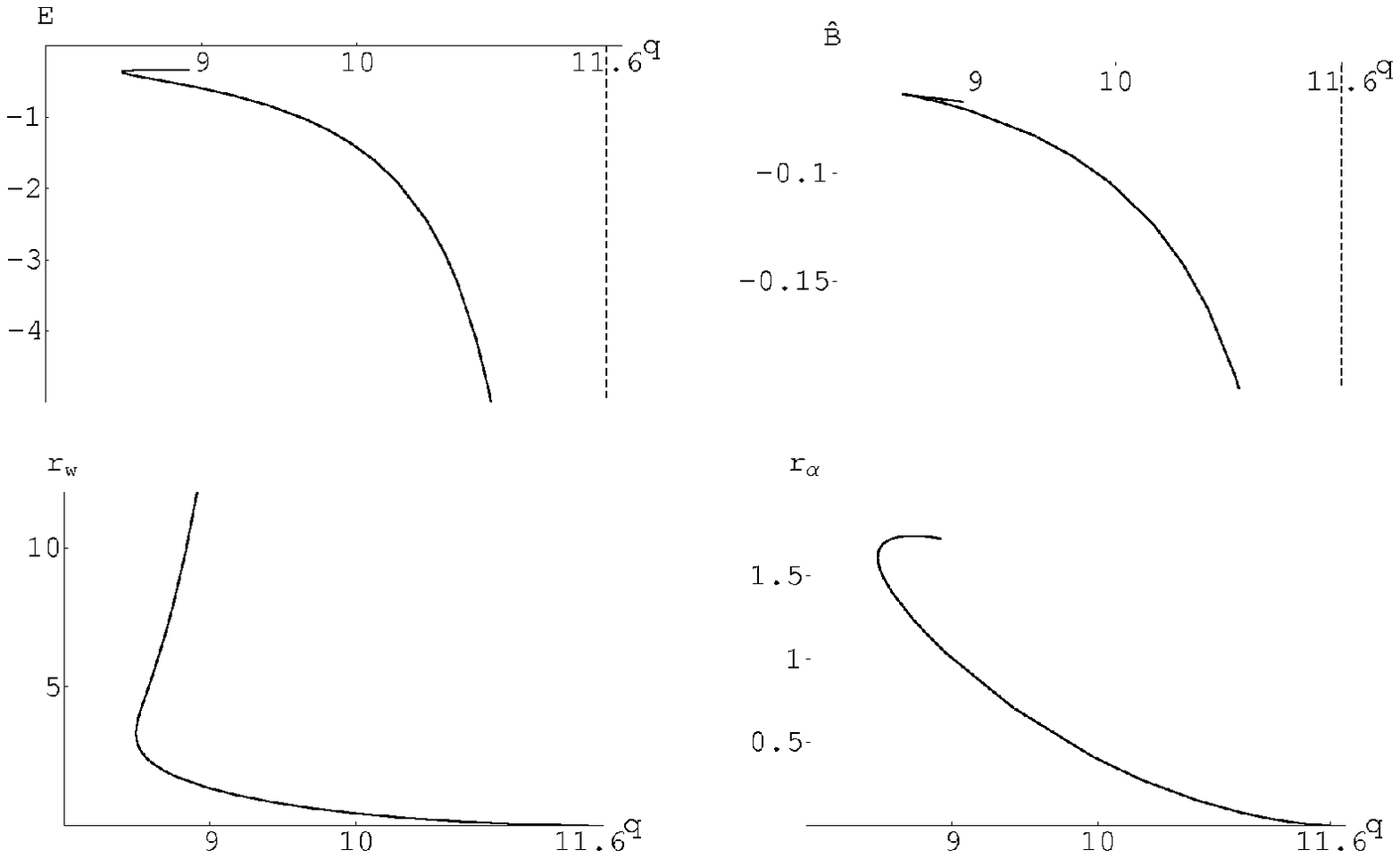}}
\caption{Reciprocal coupling limit:
The energy spectrum and characteristic radii for the ground state}
	\label{limit3}
\end{figure}

\begin{figure}[thb]
	\centerline{\epsfbox{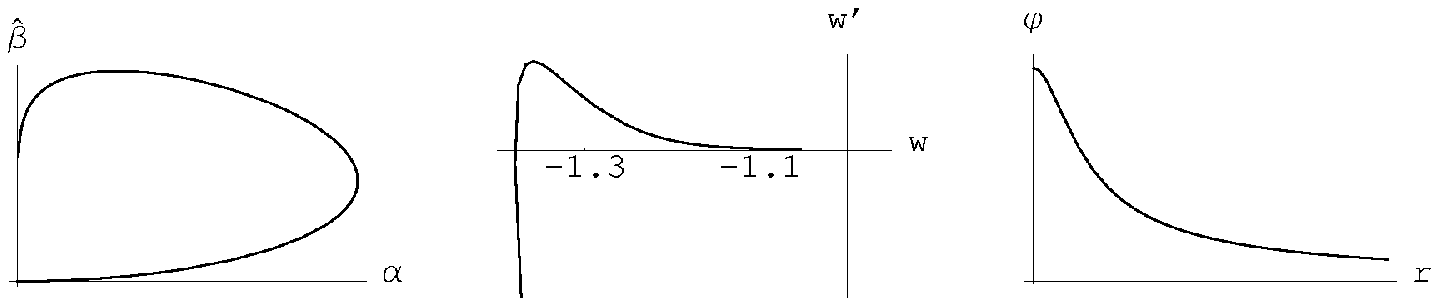}}
	\caption{Reciprocal coupling limit:
The ground state for $q=11.5838$, $E=512161$.}
	\label{limit5}
\centerline{} \centerline{}
	\centerline{\epsfbox{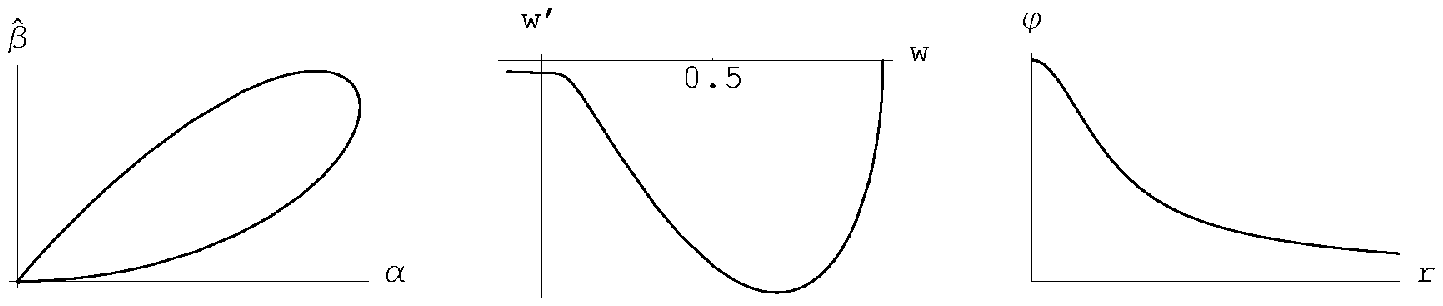}}
	\caption{Reciprocal coupling limit:
The ground state for $q=8.76701$, $E=0.334974$.}
	\label{limit6}
\centerline{} \centerline{}
	\centerline{\epsfbox{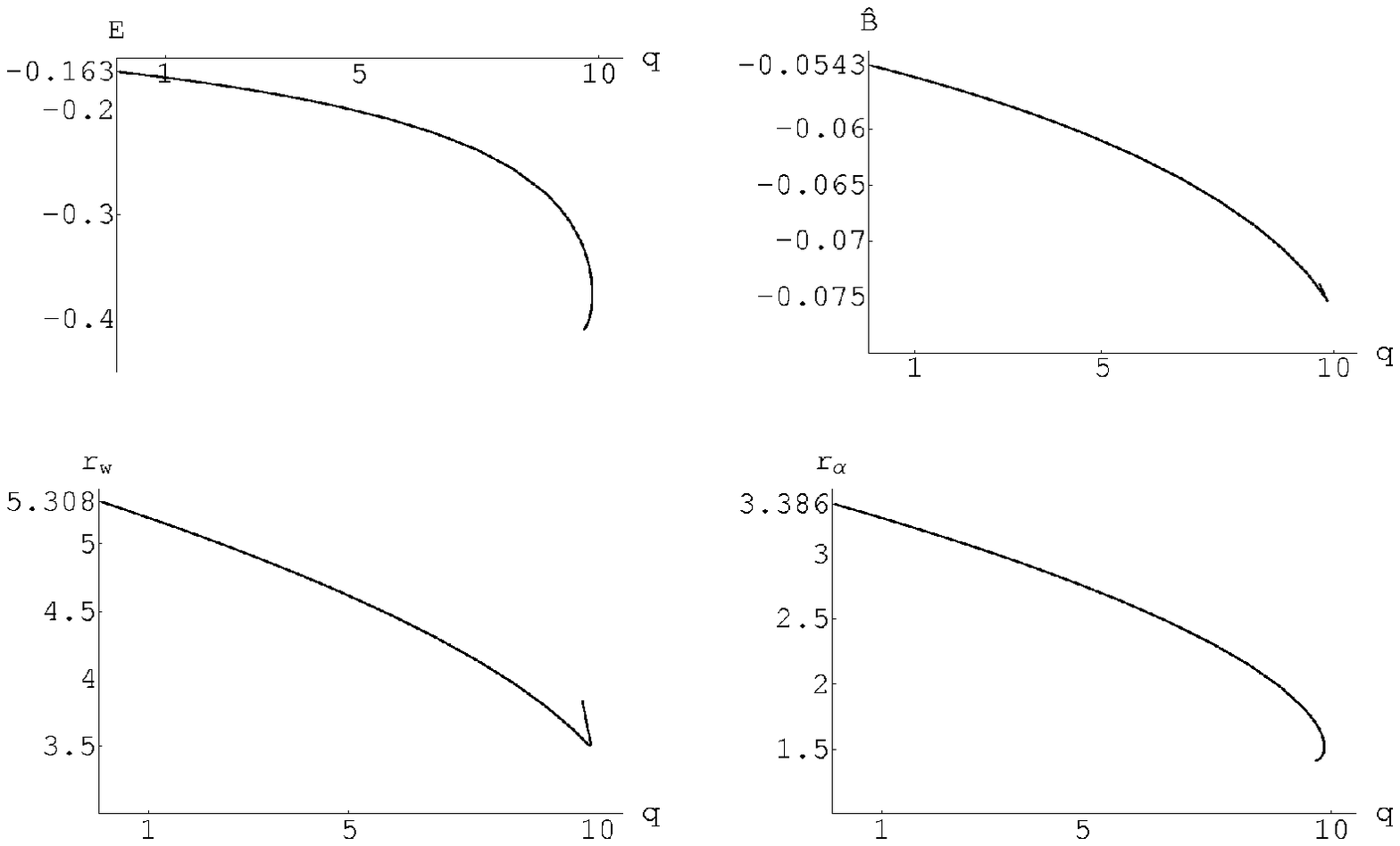}}
\caption{Reciprocal coupling limit:
The energy spectrum and characteristic radii for the first excited state}
	\label{limit7}
\end{figure}

\begin{figure}[thb]
	\centerline{\epsfbox{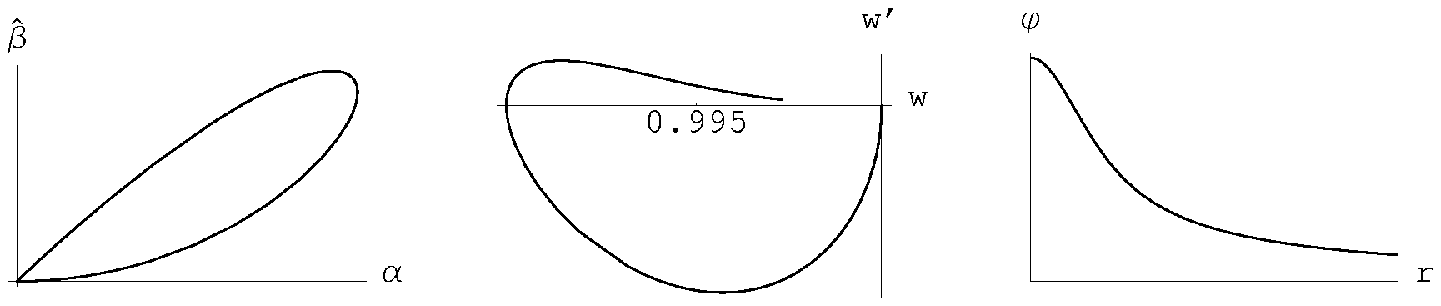}}
	\caption{Reciprocal coupling limit:
The first excited state for $q=0.285092$, $E=0.16431$.}
	\label{limit8}
\centerline{} \centerline{}
	\centerline{\epsfbox{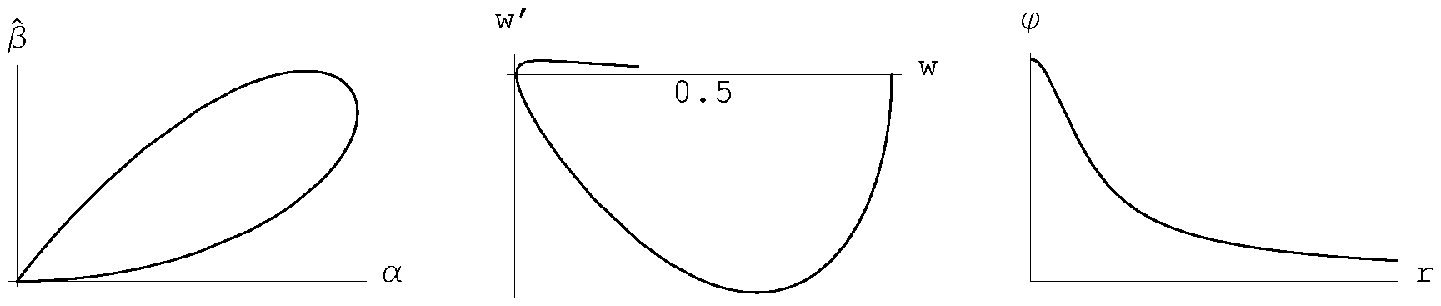}}
	\caption{Reciprocal coupling limit:
The first excited state for $q=9.75946$, $E=0.405186$.}
	\label{limit9}
\centerline{} \centerline{}
	\centerline{\epsfbox{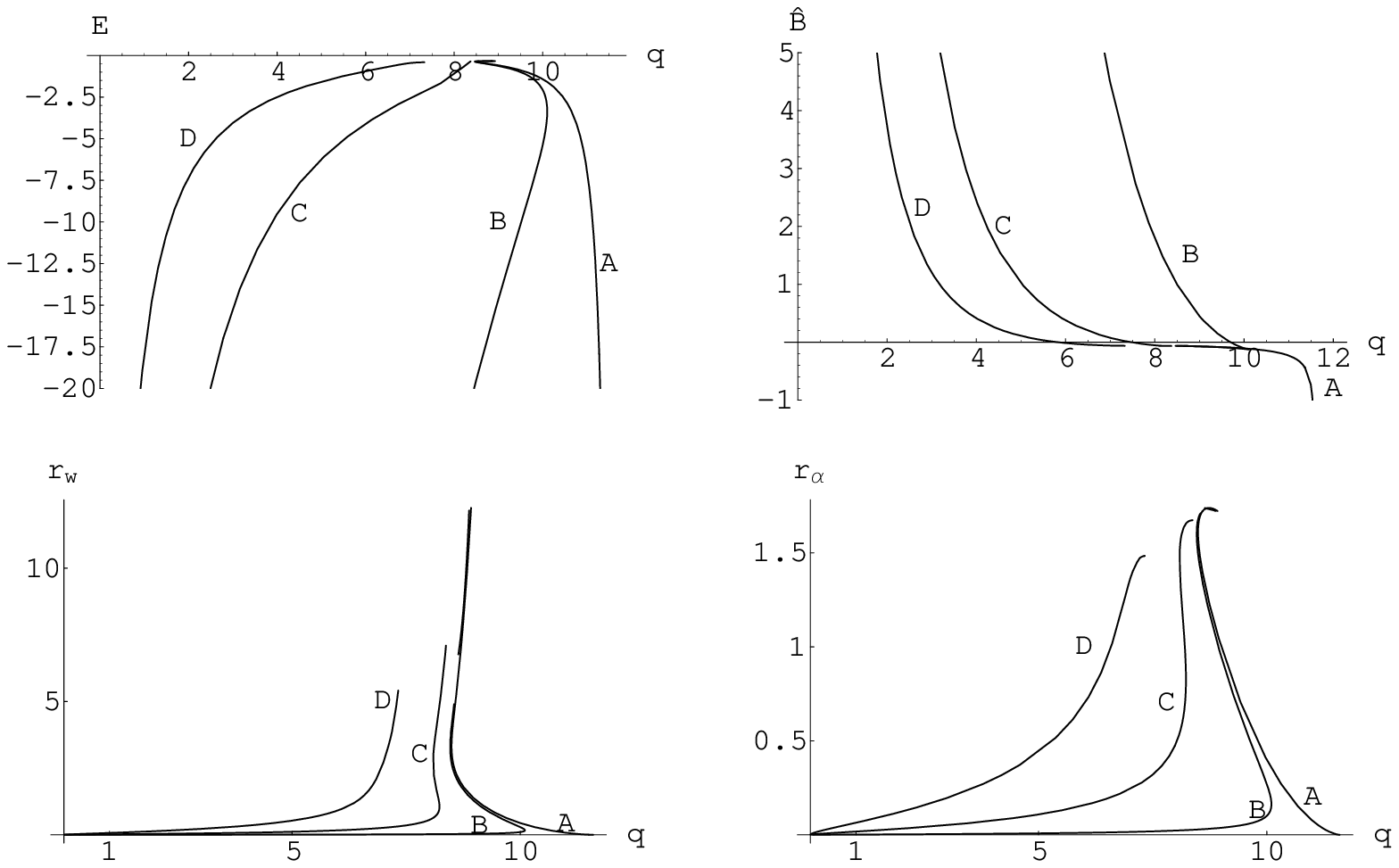}}
\caption{The energy spectrum and characteristic radii for the EDYM ground
state and $\varepsilon=0$ (A), $\varepsilon=0.0003933$ (B),
$\varepsilon=0.005322$ (C), and $\varepsilon=0.02067$ (D).}
	\label{EDYM1}
\end{figure}

\begin{figure}[t]
	\centerline{\epsfbox{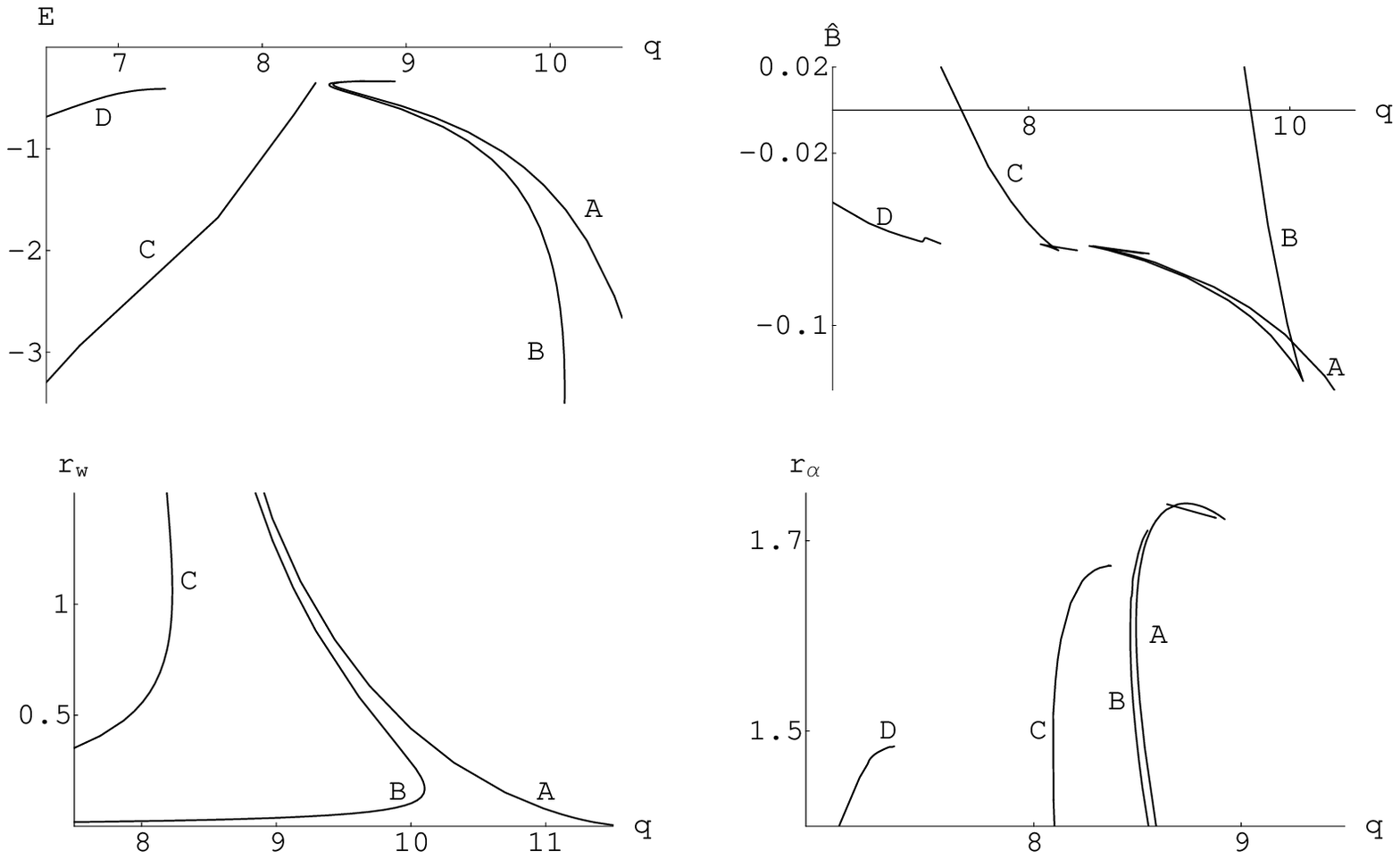}}
\caption{Details of the energy spectrum and characteristic radii for the
EDYM ground
state and $\varepsilon=0$ (A), $\varepsilon=0.0003933$ (B),
$\varepsilon=0.005322$ (C), and $\varepsilon=0.02067$ (D).}
	\label{EDYM2}
\end{figure}

\begin{figure}[thb]
	\centerline{\epsfbox{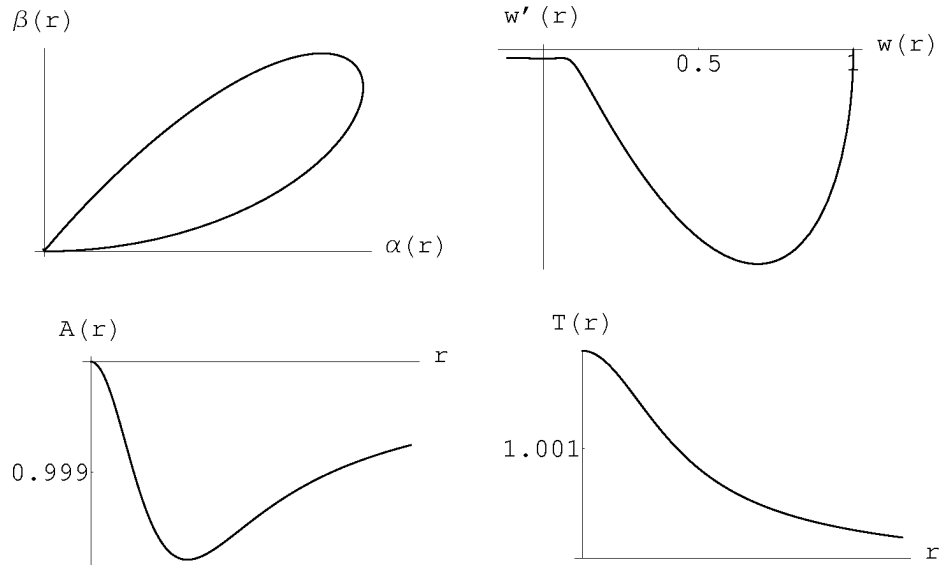}}
\caption{The EDYM ground state for $q=8.8373$, $E=0.3367$, and
$\varepsilon=0.0003933$.}
	\label{EDYM3}
\end{figure}

\begin{figure}[thb]
	\centerline{\epsfbox{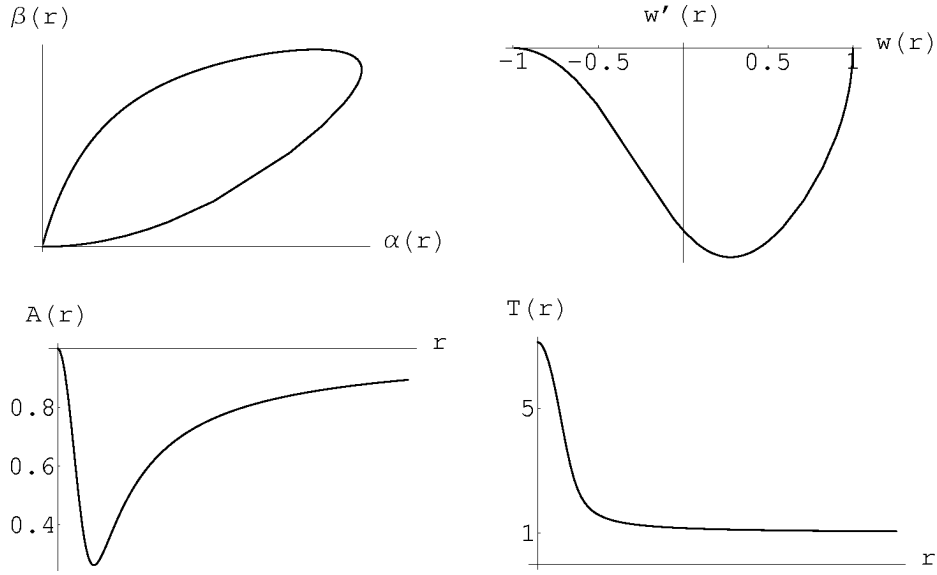}}
\caption{The EDYM ground state for $q=0.005881$, $E=17867.5$, and
$\varepsilon=0.005322$.}
	\label{EDYM4}
\end{figure}

We now describe the solutions we found. Exactly as for the 
reciprocal limit system in Section~\ref{sec3}, we restricted attention to 
solutions with rotation number zero for the spinors and a rotation 
angle of $\pi$ or $2 \pi$ for the YM potential. We again refer to 
these types of solutions as the ``ground state'' and the ``first 
excited state,'' respectively. For the ground state solutions, the 
energy spectrum and the characteristic radii are in 
Figures~\ref{EDYM1} and~\ref{EDYM2} plotted 
for different values of the parameter $\varepsilon$ (the 
characteristic radii are again defined 
by~(\ref{eq:rw1})). The curves A for $\varepsilon=0$ coincide with the 
plots for the reciprocal limit system in Figure~\ref{limit3}. For 
small values of the parameter $\varepsilon$, there are solutions for
the EDYM equations which are close to the solutions of the reciprocal limit 
equations (compare the curves A and B in Figure~\ref{EDYM1}). In this 
parameter regime, the EDYM solutions look typically as shown in 
Figure~\ref{EDYM3}.
The metric functions $A$ and $T$ are both close to one; thus the 
gravitational interaction is weak, in agreement with our 
considerations after (\ref{eq:2A}). The spinors and the YM potential 
look very similar to the solution of the reciprocal limit equations in 
Figure~\ref{limit6}. {\em{We conclude that the reciprocal limit system of 
Section~\ref{sec3} indeed describes a significant limiting case of the 
EDYM equations.}}
However, one also sees that even for small $\varepsilon$, not all the 
solutions of the EDYM equations are close to the reciprocal limit 
solutions. 
More precisely, curve B leaves the vicinity of curve A at $q \approx 
10$ (see Figure~\ref{EDYM2}). If one follows curve B after it branches off
from curve A, the parameter $q$ first increases up to a turning point, and
then decreases to $q=0$. If $\varepsilon$ gets large, the solution curves 
no longer come so close to the reciprocal limit solutions (see curves C 
and D). The maximum of $q$ decreases (see curve C) and finally 
disappears (see curve D).
Figure~\ref{EDYM4} shows a typical solution for small $q$.
We note that in this parameter region, the metric functions $A$ and $T$ are
not near one; this explains why the reciprocal limit 
equations are no longer a good approximation. Indeed, the 
potentials $w$, $A$, and $T$ now resemble a BM solution of 
the EYM equations \cite{BM}, and the spinors look like the solution 
of the Dirac equation in the BM background. Hence $q \rightarrow 0$ 
corresponds to the limit of weakly coupled spinors; i.e. spinors in a
fixed BM background. Notice that the 
characteristic radii go to zero and the energies go to infinity in the 
limit $q \rightarrow 0$ (see Figure~\ref{EDYM1}).
This can be understood from our rescalings. 
Namely, for the (unscaled) solutions of our modified EDYM system, the 
BM solutions are easily obtained by taking the limit $\alpha_1 \rightarrow 
0$ (in which the spinors go uniformly in $r$ to zero). In this limit, the 
normalization integral (\ref{eq:cn1}) tends to zero, and thus the 
rescalings (\ref{eq:res0})--(\ref{eq:res}) lead to a singular 
behavior of the rescaled solutions for $q \rightarrow 0$. To 
summarize, there is a one-parameter family of solutions (obtained by
continuously changing the coupling constants), connecting the BM solutions
to our reciprocal limit solutions

We remark that our plots of the curves B have a small gap at $q 
\approx 8.7$. The reason is that in this region the numerics become 
unstable, and could not be carried out with our methods. But we were 
able to construct two branches of solutions which approach the 
problematic region from both sides. We suspect that the instability 
of the numerics is merely an artifact of our rescaling method, but it 
might well be an indication for a possible bifurcation point in this 
region. For the other curves C and D, we analyzed only the branch of 
solutions which extends towards smaller values of $q$.

\begin{figure}[t]
	\centerline{\epsfbox{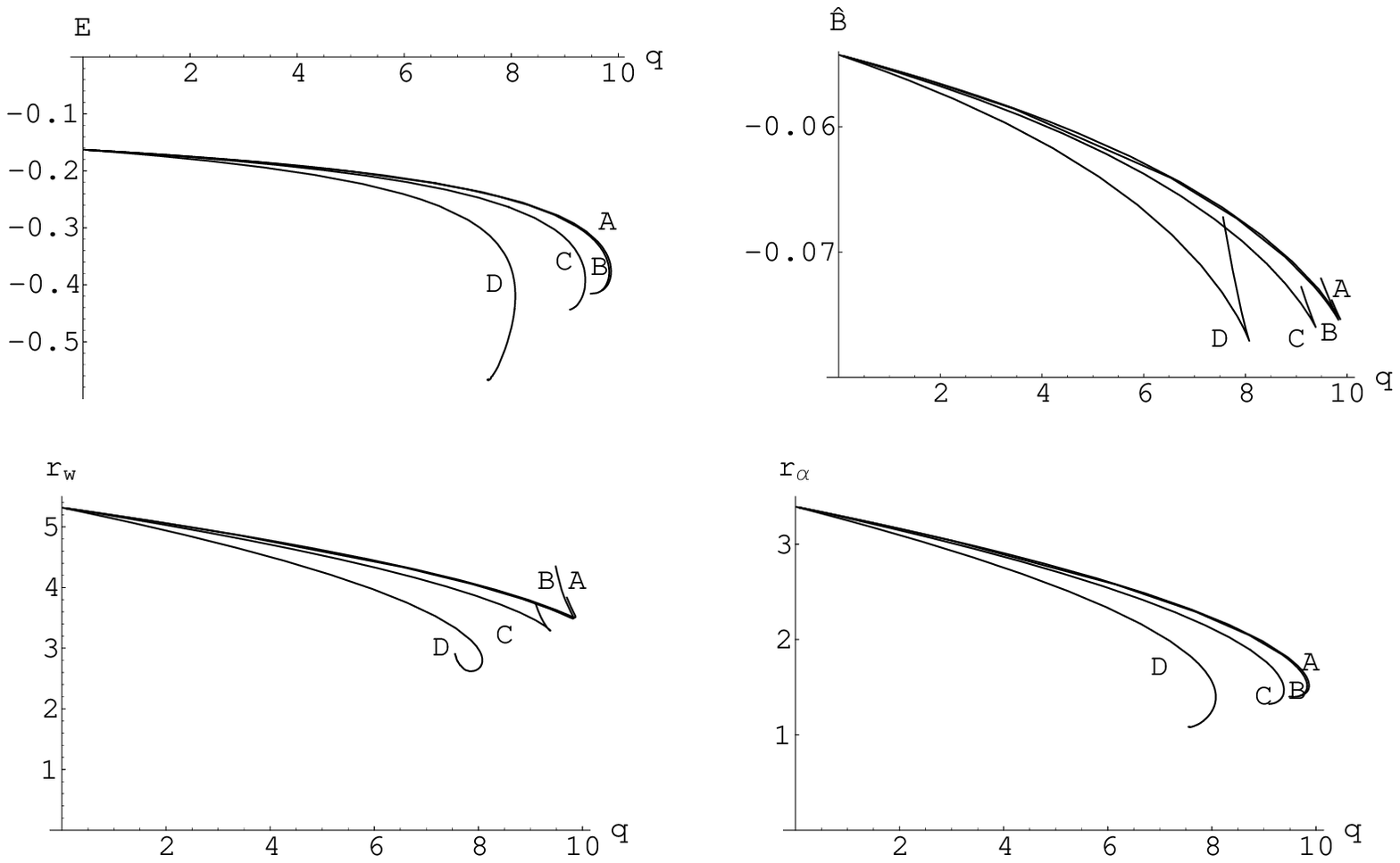}}
\caption{The energy spectrum and characteristic radii for the first excited
EDYM state and $\varepsilon=0$ (A), $\varepsilon=0.0003933$ (B),
$\varepsilon=0.005322$ (C), and $\varepsilon=0.02067$ (D).}
	\label{EDYM5}
\end{figure}

\begin{figure}[thb]
	\centerline{\epsfbox{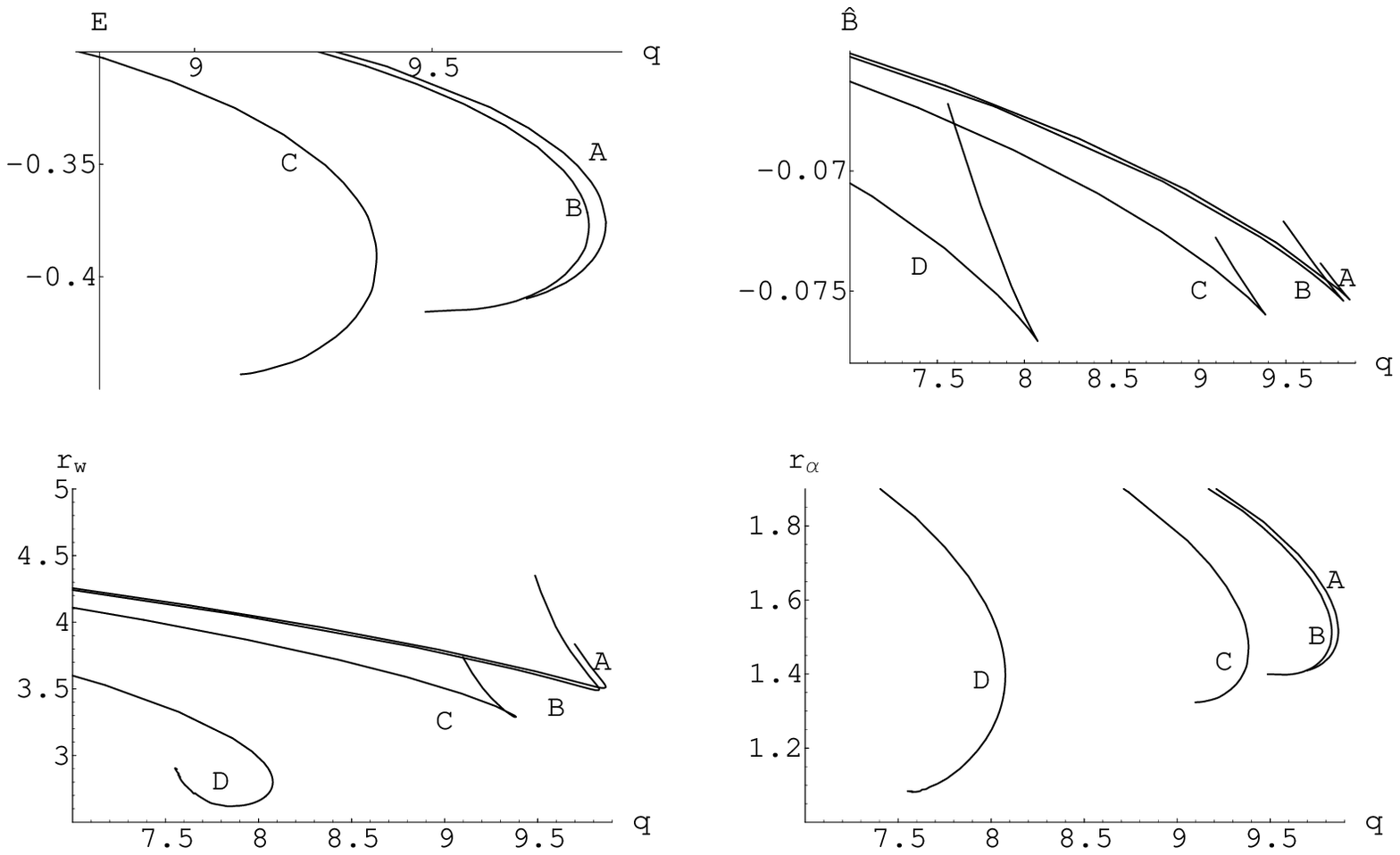}}
\caption{Details of the energy spectrum and characteristic radii for the
first excited EDYM state and $\varepsilon=0$ (A), $\varepsilon=0.0003933$
(B), $\varepsilon=0.005322$ (C), and $\varepsilon=0.02067$ (D).}
	\label{EDYM6}
\end{figure}

\begin{figure}[thb]
	\centerline{\epsfbox{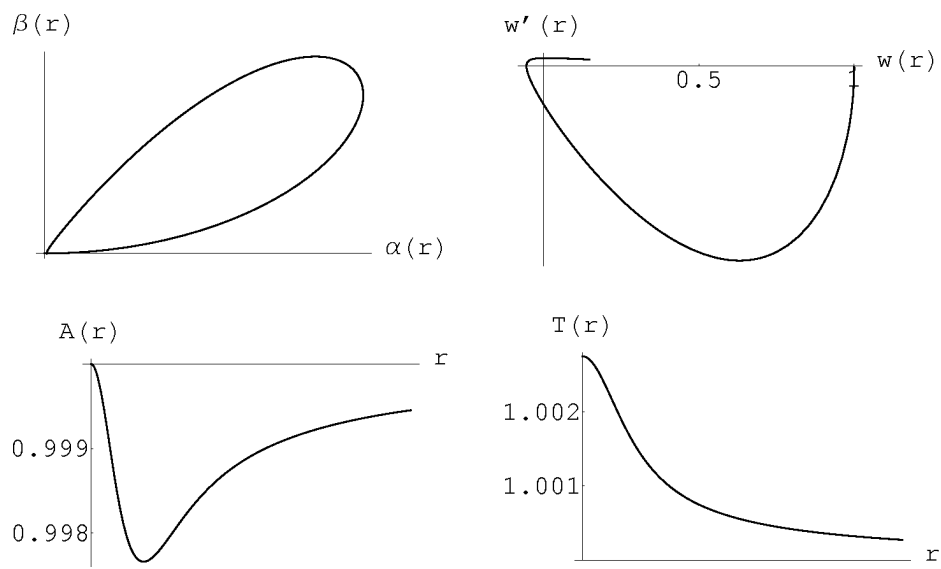}}
\caption{The first excited EDYM state for $q=9.548$, $E=0.4153$, and
$\varepsilon=0.0003933$.}
	\label{EDYM7}
\end{figure}

\begin{figure}[thb]
	\centerline{\epsfbox{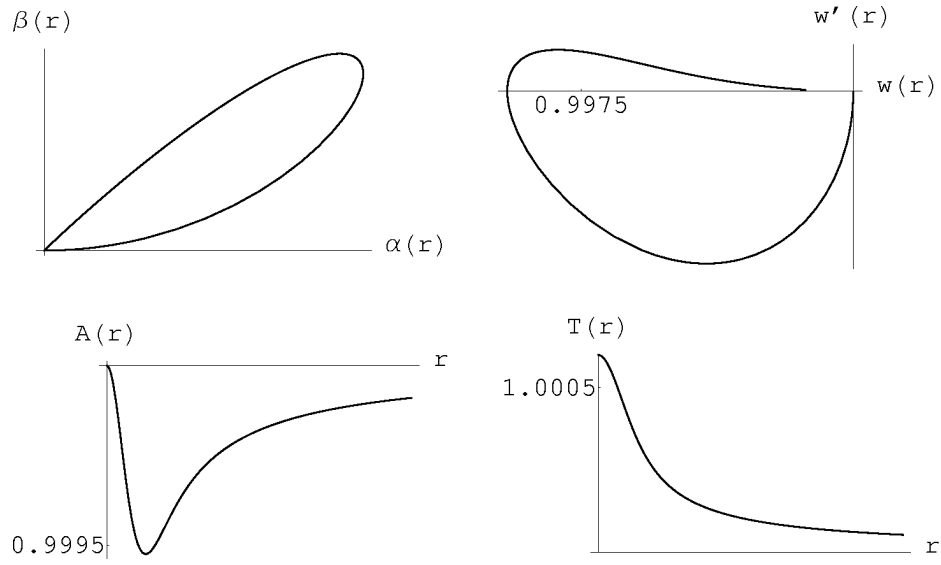}}
\caption{The first excited EDYM state for $q=0.09013$, $E=0.1634$, and
$\varepsilon=0.02067$.}
	\label{EDYM8}
\end{figure}

For the first excited state, the energy spectrum and characteristic 
radii are plotted in Figures~\ref{EDYM5} and~\ref{EDYM6}
(the characteristic radii are again defined by (\ref{eq:rw2})). The 
curves A for $\varepsilon=0$ correspond to the solutions of the 
reciprocal limit equations in Figure~\ref{limit7}. In contrast to the 
situation for the ground state, the solutions for small $\varepsilon$ 
are all close to the reciprocal limit solutions (compare the curves A and B). 
Figure~\ref{EDYM7} shows a typical solution for large $q$; one sees 
that the spinors and YM potential look similar to those in Figure~\ref{limit9}.
The form of the energy spectrum and the characteristic radii 
gradually change when $\varepsilon$ is increased; for example, the cusp in 
the $(q, r_w)$-plot becomes smooth (see curve D). It is 
interesting that for $q \rightarrow 0$, the curves converge 
independent of $\varepsilon$ to a single limit point (see Figure~\ref{EDYM5}).
This limit point was 
already described at the end of Section~\ref{sec3} as the case when 
the spinors form a bound state due to their gravitational attraction, 
and the spinors generate a YM field (see (\ref{eq:DN1})--(\ref{eq:DN4})).
This picture is in agreement with our numerics, since the 
spinors and metric functions, for a solution near this limit point, look 
similar to the ED solutions \cite{FSY1} in the Newtonian limit, and 
$w \approx 1$ (see Figure~\ref{EDYM8}).
The fact that this limit point is independent of $\varepsilon$ follows, 
because as explained at the end of Section~\ref{sec3}, for $q 
\rightarrow 0$, the YM equation decouples from the ED equations.
For clarity, we point out that it would not be correct to say that 
the gravitational interaction dominates the YM interaction in the 
limit $q \rightarrow 0$. Namely, according to (\ref{eq:cp1}), the 
ratio of the gravitational and YM coupling constants is kept fixed, 
and thus $q \rightarrow 0$ corresponds to the limit where both 
coupling constants go to zero at the same rate. Nevertheless,
the YM field has for $q \rightarrow 0$ no influence on the energy spectrum
and the characteristic radii.

\begin{figure}[thb]
	\centerline{\epsfbox{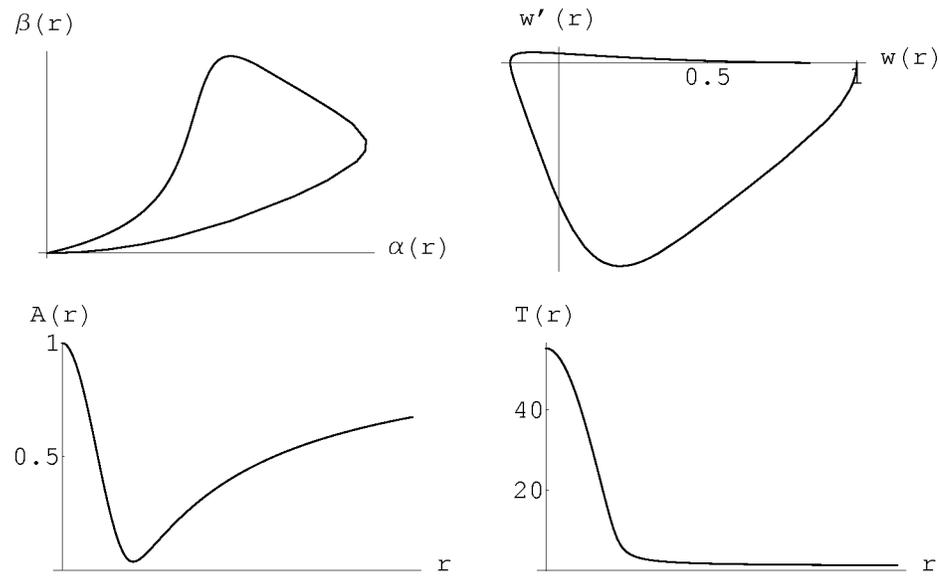}}
\caption{The deformation of the first excited BM state for
$q=0.00047737$, $E=367616$, and $\varepsilon=0.02067$.}
	\label{EDYM9}
\end{figure}

\begin{figure}[t]
	\centerline{\epsfbox{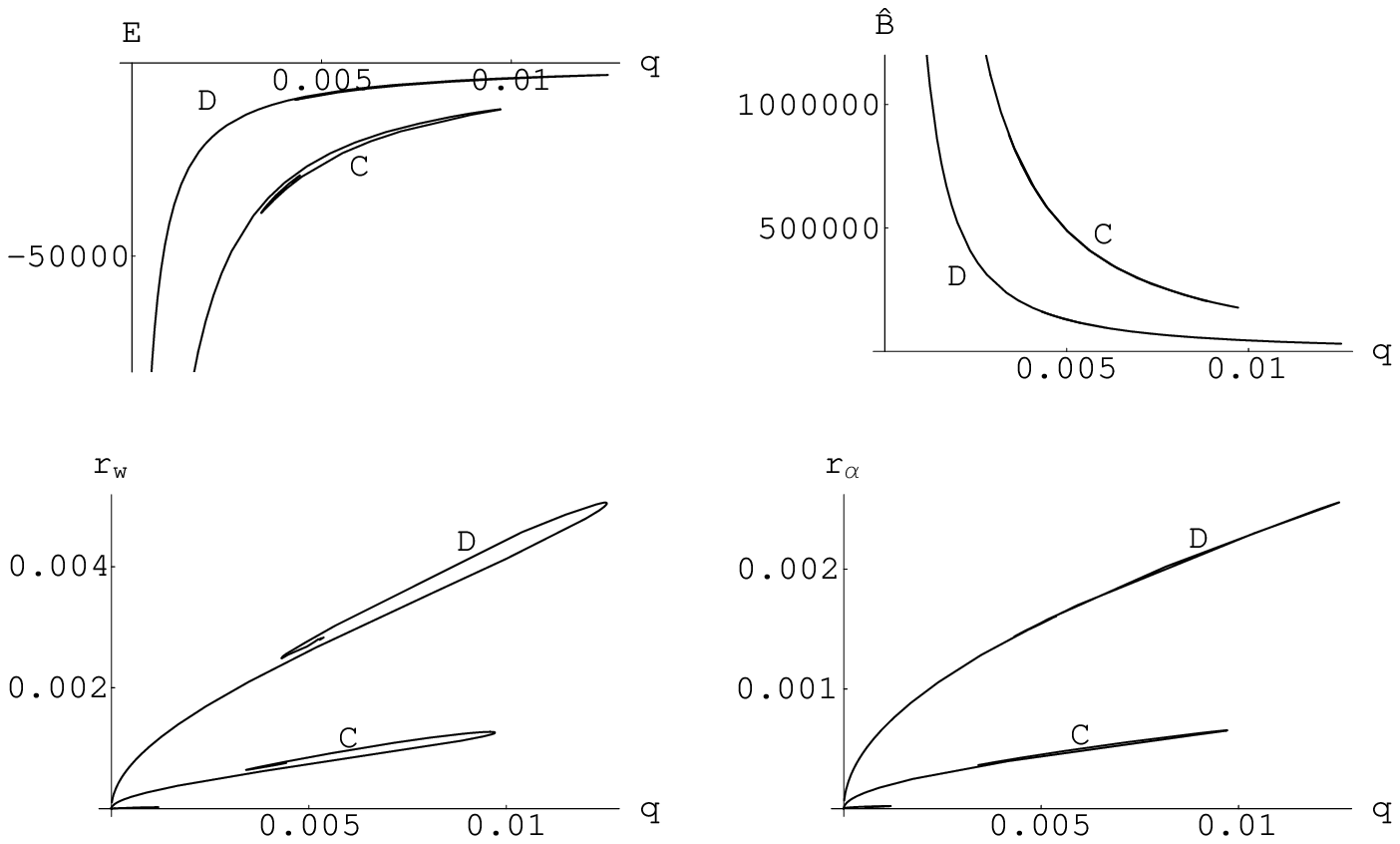}}
\caption{The energy spectrum and characteristic radii for the deformation of
the first excited BM state for
$\varepsilon=0.005322$ (C) and $\varepsilon=0.02067$ (D).}
	\label{EDYM10}
\centerline{} \centerline{}
	\centerline{\epsfbox{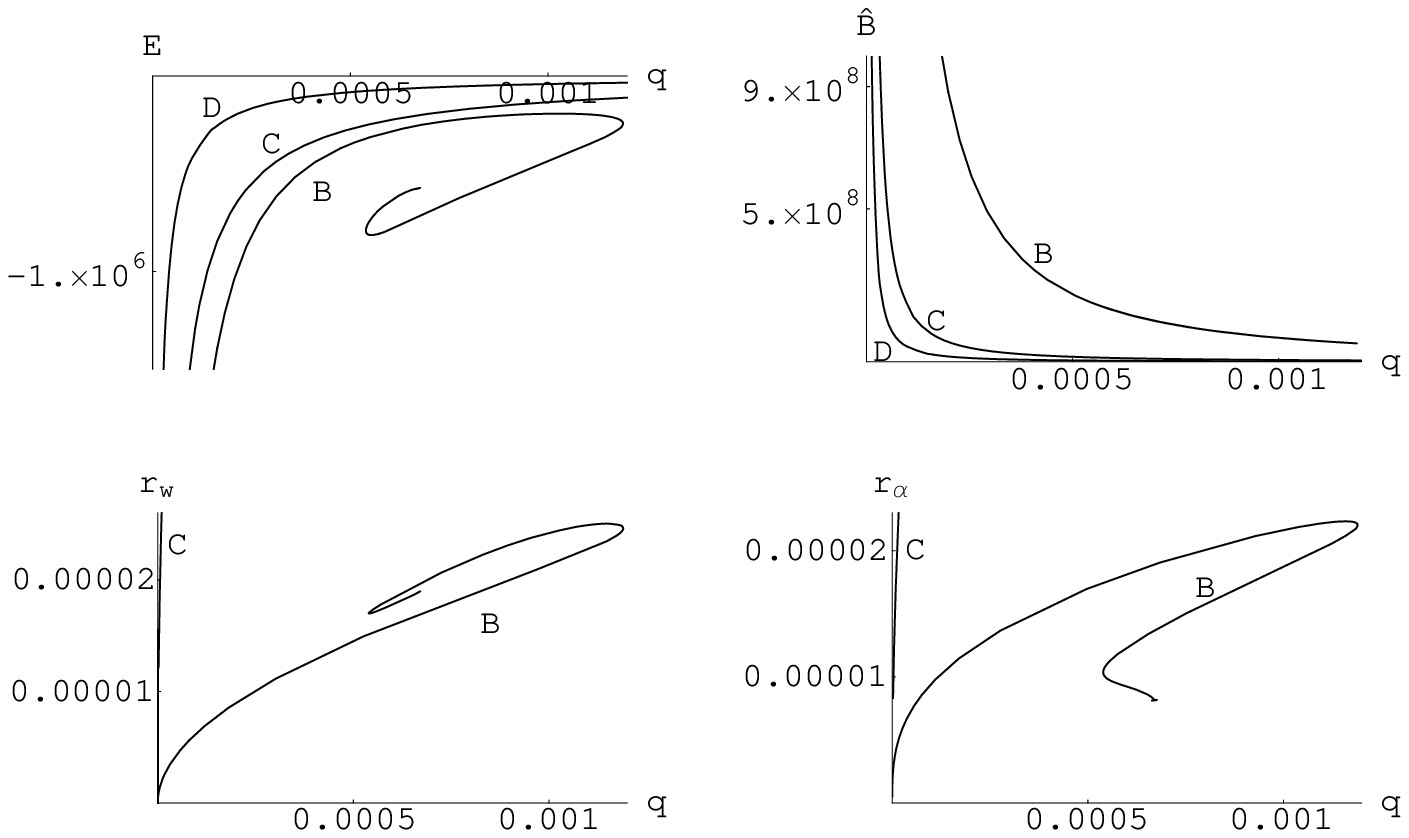}}
\caption{The energy spectrum and characteristic radii for the deformation of
the first excited BM state for $\varepsilon=0.0003933$ (B),
$\varepsilon=0.005322$ (C), and $\varepsilon=0.02067$ (D).}
	\label{EDYM11}
\end{figure}

\begin{figure}[thb]
	\centerline{\epsfbox{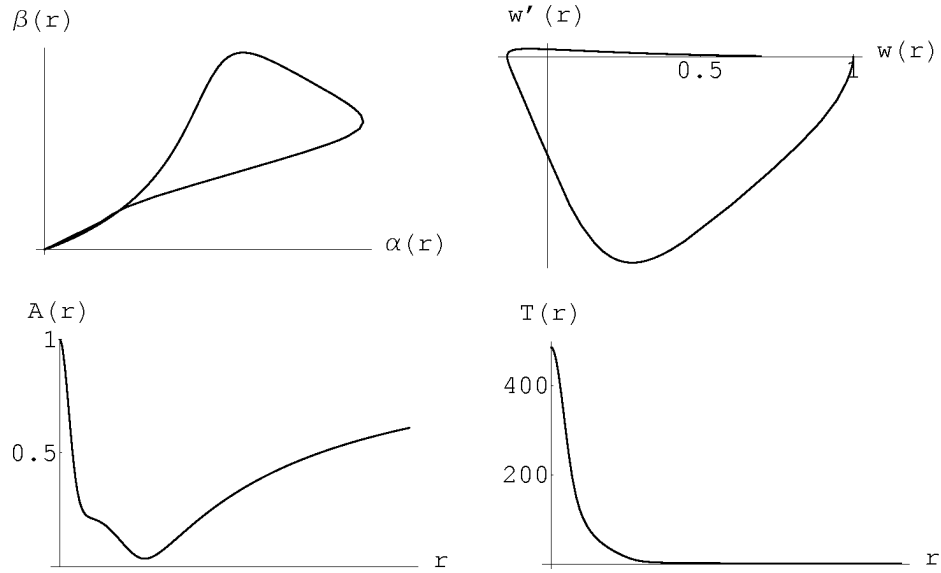}}
\caption{The deformation of the first excited BM state for
$q=0.0005528$, $E=729921$, and $\varepsilon=0.0003933$.}
	\label{EDYM13}
\end{figure}

A main qualitative difference between the ground state and the first 
excited state is that for the first excited state, we could not 
continuously join the solutions of the reciprocal limit equations with 
a BM solution. In order to see how this comes about, we did 
numerical calculations starting with a Dirac particle in the BM 
background (similar to that shown in Figure~\ref{EDYM9})
and gradually increased the coupling of the spinors to gravity and to
the YM field. For these ``deformations of the first excited 
BM state,'' the curves of the energy 
spectrum and the characteristic radii have spirals, whose size and 
shape drastically changes when $\varepsilon$ is increased, see 
Figures~\ref{EDYM10} and~\ref{EDYM11}.
In the parameter regime where the energy plots spiral around, the 
spinors have self-intersections similar as observed for the ED 
solutions~\cite{FSY1}, see Figure~\ref{EDYM13}.

\begin{figure}[thb]
	\centerline{\epsfbox{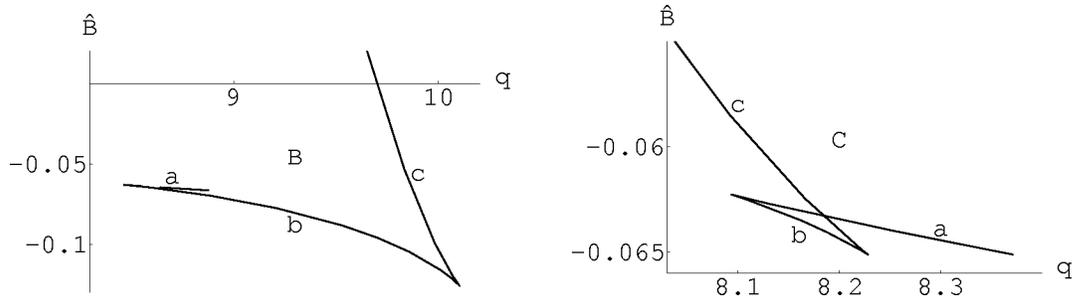}}
\caption{Detailed plots of the total binding energy of the EDYM ground
state for $\varepsilon=0.0003933$ (B) and $\varepsilon=0.005322$ (C).}
	\label{EDYM14}
\end{figure}

We now discuss the stability of our solutions. The relevant parameter
for the stability analysis is the total binding energy $\hat{B}$. Namely,
if $\hat{B}$ is negative and smaller than the total energies of all other
states, then energy is needed to break up the binding or to make a
transition to any of the other states, and therefore for
physical reasons the solution must be stable. Clearly, this energy
argument does not provide a rigorous stability proof, and it
also cannot replace the numerical analysis of linear stability (like e.g.
in \cite{SZ} or \cite[Section~8]{FSY1}), but it gives a strong
indication for stability and is therefore commonly used (see e.g.~\cite{L}).
Let us first apply this energy argument to the ground state solutions of
Figures~\ref{EDYM1} and~\ref{EDYM2}. One sees that the total energy
becomes negative for large $q$. For the curves B and C, this region
is plotted in more detail in Figure~\ref{EDYM14}.
For the solutions on branch b, the total binding energy is minimal,
and thus this branch is stable.
Applying Conley index methods with $q$ as the bifurcation parameter (see
\cite{Sm}), we obtain, as in \cite{FSY1}, that the two other branches
a and c are unstable. Indeed, the instability of branch c follows
also from the continuity of the Conley index and the fact that in
the limit $q \rightarrow 0$, this branch goes over to the ground state BM
solution which is known to be unstable \cite{SZ}.
When $\varepsilon$ is increased (see curve D, Figures~\ref{EDYM1}
and~\ref{EDYM2}), only one branch of
solutions remains, which comprises the BM solutions as a limiting case and is
therefore unstable. More precisely, the one-parameter family has in this
case no bifurcation points, and in the limit $q \rightarrow 0$ the
solutions tend to an unstable BM solution. Thus using Conley index
techniques, it follows that the entire one-parameter family is unstable.
We conclude that {\em{for small $\varepsilon$, there is
a stable branch of ground state solutions for which $q$ lies in a finite
interval away from $q=0$}}; all other ground state solutions are unstable.
For the stability of the first excited state, we consider the plots of
Figures~\ref{EDYM5} and~\ref{EDYM6}. Since for $q \rightarrow 0$, the
spinors and metric functions go over to the Newtonian limit of the ED
solutions, we conclude from~\cite{FSY1} that {\em{the branch of solutions
starting at $q=0$ should be stable}}. This is in agreement with our above
energy argument, because on this branch the total binding energy $\hat{B}$
is negative, and is smaller than the
total binding energy of the second branch of solutions, which comes out of
the bifurcation point located at the maximum of $q$.
Again, Conley index theory yields that this second branch is unstable.
For the deformations of the first excited BM state, the total binding
energy is positive (see Figures~\ref{EDYM11} and \ref{EDYM13}), and hence
these solutions should be unstable. Indeed, for the branch of solutions which
extends up to $q=0$ (i.e.\ before the first bifurcation point), this also
follows from the continuity of the Conley index and the instability of the
first excited BM solution.

\newpage
\clearpage

\clearpage

\clearpage

\clearpage

\clearpage


\begin{footnotesize}
\begin{tabular}{lclcl}
\\
Felix Finster & $\;$ & Joel Smoller & $\;$ & Shing-Tung Yau \\
Max Planck Institute MIS && Mathematics Department && Mathematics 
Department \\
Inselstr.\ 22-26 && The University of Michigan && Harvard University \\
04103 Leipzig, Germany && Ann Arbor, MI 48109, USA
&& Cambridge, MA 02138, USA \\
{\tt{Felix.Finster@mis.mpg.de}} && {\tt{smoller@umich.edu}}
&& {\tt{yau@math.harvard.edu}}
\end{tabular}
\end{footnotesize}


\clearpage
\newpage
\begin{thebibliography}{99}
\bibitem{BM} Bartnik, R., and McKinnon, J., ``Particlelike solutions of the
Einstein-Yang-Mills equations,'' {\em{Phys. Rev. Lett.}}\ 61 (1988)
141-144
\bibitem{SZ} Straumann, N,. and Zhou, Z., ``Instability of the
Bartnik-McKinnon solution,'' {\em{Phys. Lett. B}}\ 237 (1990) 353-356
\bibitem{FSY1} Finster, F., Smoller, J., and Yau, S.-T., 
``Particlelike solutions of the Einstein-Dirac equations,''
gr-qc/9801079, {\em{Phys.\ Rev.}}\ D 59 (1999) 104020
\bibitem{FSY2} Finster, F., Smoller, J., and Yau, S.-T., 
``Particlelike solutions of the Einstein-Dirac-Maxwell equations,''
gr-qc/9802012, {\em{Phys.\ Lett.}}\ A 259 (1999) 431-436
\bibitem{LL} Landau, L.D., Lifshitz, E.M., ``Quantum Mechanics,'' 
Pergamon Press (1977)
\bibitem{FSY3} F.\ Finster, J.\ Smoller, and S.-T.\ Yau, ``Non-Existence of
Time-Periodic Solutions of the Dirac Equation in a Reissner-Nordstr\"om Black
Hole Background,'' gr-qc/9805050, {\em{J.\ Math.\ Phys.}} 41 (2000) 2173-2194
\bibitem{Y} Yang, C.N., and Wu, T.T., ``Some solutions of the 
classical isotopic gauge field equations,'' in ``Properties of Matter 
Under Unusual Conditions,'' H.\ Mark and S. Fernbach eds., New York, 
Wiley Interscience (1969) 349-354
\bibitem{S} Stoer, J., and Bulirsch, R., ``Numerische Mathematik 2,''
3rd edition, Springer (1990)
\bibitem{L} Lee, T.D., ``Mini-soliton stars,'' {\em{Phys.\ Rev.\ D}} 25 (1987)
3640-3657
\bibitem{Sm} Smoller, J., ``Shock Waves and Reaction-Diffusion 
Equations,'' 2nd ed., Springer (1994)
\end{thebibliography}
\end{document}